\documentclass[useAMS,usenatbib]{mn2e}

\title[Why are most molecular clouds not gravitationally bound?]
{Why are most molecular clouds not gravitationally bound?}
\author[C. L. Dobbs, A. Burkert and J. E. Pringle]
{C. L. Dobbs$^{1,2}$\thanks{E-mail:
cdobbs@mpe.mpg.de}, A. Burkert$^{2,3}$ and J. E. Pringle$^{4}$\\
$^1$ Max-Planck-Institut f\"ur extraterrestrische Physik, Giessenbachstra\ss{}e, D-85748 Garching, Germany \\
$^2$ Universitats-Sternwarte M\"unchen, Scheinerstra\ss{}e 1, D-81679
M\"unchen, Germany \\
$^3$ Max-Plack fellow, Max-Planck-Institut f\"ur extraterrestrische Physik, Giessenbachstra\ss{}e, D-85748 Garching, Germany \\
$^4$ Institute of Astronomy, Madingley Road, Cambridge, CB3 0HA}
\usepackage{amssymb}
\usepackage{amsmath}
\usepackage{graphicx}
\usepackage{epsfig}
\usepackage{multirow}

\begin{document}
\date{\today}

\pagerange{\pageref{firstpage}--\pageref{lastpage}} \pubyear{0000}

\maketitle

\label{firstpage}
\begin{abstract}
The most recent observational evidence seems to indicate that giant
molecular clouds are predominantly gravitationally unbound
objects. In this paper we show that this is a natural consequence of a
scenario in which cloud-cloud collisions and stellar feedback regulate
the internal velocity dispersion of the gas, and so prevent global
gravitational forces from becoming dominant. Thus, while the molecular
gas is for the most part gravitationally unbound, local regions within
the denser parts of the gas (within the clouds) do become bound and
are able to form stars. We find that the observations, in terms of
distributions of virial parameters and cloud structures, can be well
modelled provided that the star formation efficiency in these bound
regions is of order 5 -- 10 percent. We also find that in this picture
the constituent gas of individual molecular clouds changes over relatively short
time scales, typically a few Myr.
\end{abstract}

\section{Introduction}
The belief that molecular clouds are gravitationally bound objects
has led to a long-standing problem in star formation. If clouds are
bound, and if they collapse in of order a free-fall time, then the
rate of star formation should be around two orders of magnitude
greater than what is observed \citep{Zuckerman1974}. To circumvent
this there have been two main approaches. Firstly, one can assume that
molecular clouds undergo collapse on longer
timescales, requiring that some process prevents, or at least slows their collapse.
This could be magnetic fields and slow ambipolar diffusion
\citep{Shu1987,Basu1994,Allen2000,Mous2006,Shu2007}, or microturbulence
\citep{Krum2005,Krumholz2006}. {However 
turbulence can also promote collapse on small scales
\citep{Klessen2000,MacLow2004}, 
(including MHD turbulence \citealt{Heitsch2001,Vaz2005,Elmegreen2007}).
Alternatively, in light of recent observational evidence, one can assume that
clouds are typically short-lived entities \citep{Hart2001,Elmegreen2002,
Elmegreen2007, Ball2007} and posit
mechanisms which prevent most of the gas forming stars,
such as stellar feedback and magnetic fields \citep{Vaz2005,Elmegreen2007,Price2008}. 
However if the cloud is not globally bound
in the first place, as suggested by recent observational evidence, we
already ameliorate the disparity between the existence of 
short-lived molecular clouds and the global star formation rate.

The virial parameter of a molecular cloud is usually defined as 
\begin{equation}
\alpha=\frac{5 \sigma_v^2 R}{G M}
\end{equation}
 (e.g. \citealt{Bertoldi1992,Dib2007}), where $\sigma_v$ is the
line of sight velocity dispersion and R is a measure of the radius of
the cloud.  If the cloud is in virial equilibrium, then $\alpha=1$ and
$2T+W$=0 where $T$ is the kinetic and $W$ the gravitational
energy, whereas $T+W=0$ corresponds to the zero energy configuration. 
The value of $\alpha$ is simply a measure of the ratio of
kinetic to gravitational energies, and finding $\alpha$ both
observationally and from simulations is highly uncertain, depending
on the mass and radius determinations, and typically does not account for
magnetic fields, surface terms or projection effects \citep{Ball2006,Dib2007,Shetty2010}. Thus the estimated value
of $\alpha$ for one cloud may not predicate its consequent evolution, but the
distribution of $\alpha$ gives an indication of the importance of
gravity in a population of clouds, and whether they are predominantly
bound or unbound. 
 
In a recent observational study, \citet{Heyer2009} published revised
estimates of molecular cloud masses, sizes and virial parameters from
the previous seminal work by \citet{Solomon1987}. Although they claim
that molecular clouds are virialised, their plots seem to indicate
that in fact most of the clouds are unbound (as also found by
\citealt{Heyer2001}).
In Fig.~1 (lower middle panel) we plot the
value of the virial parameter taken from the clouds
observed by \citet{Heyer2009}.
\citet{Heyer2009} suggest that the masses of their clouds are
underestimated by a factor of 2 due to non-LTE effects and CO
abundance variations (see also calculations by \citet{Glover2010}
and \citet{Shetty2010b})  
within a cloud, so we have
doubled their cloud masses to produce the panel in Fig.~1 (lower
middle panel). It is evident that most of the
observed clouds have virial parameters larger than unity, indicating
that most clouds are not gravitationally bound. {Even if we
  take $\alpha>2$, whereby the clouds are strictly unbound, this
  still leaves 50 per cent unbound clouds.
Similar distributions of $\alpha$ are also found for external
GMCs, including those in M31 (\citealt{Rosolowsky2007}, from a sample of
$10^{5-6}$ M$_{\odot}$ clouds), and the GMCs detected in several nearby galaxies
by \citet{Bolatto2008}.

In a recent paper \citet{Ball2010} suggest that GMCs are undergoing
 hierarchical gravitational collapse, whereby the collapse occurs on scales
 from individual cores to the whole cloud. However it is not necessary that the cloud should be globally
 gravitationally bound.
Simulations of unbound, turbulent, GMCs naturally lead to localised star formation,
rather than spread over the entire cloud \citep{Clark2005,Clark2008}. This
then naturally leads to a low star formation efficiency. Recently,
\citet{Bonnell2010} performed calculations of an unbound $10^4$
M$_{\odot}$ cloud, and showed that stellar clusters form in
bound regions of the cloud. The internal kinematics
  of these clouds could be due to cloud-cloud collisions or
  large scale flows
\citep{Bonnell2006,Dobbs2007a,Klessen2010}, 
and/or stellar feedback (e.g. \citealt{MacLow2004} and references therein).

Most numerical work has tended to focus on calculating 
the virial parameters of
clumps within giant molecular clouds \citep{Dib2007,Shetty2010},
which, since they are the sites of star formation in GMCs, are more
likely to be bound. However in recent simulations of a
galactic disk, it has been possible
to identify individual GMCs and determine their virial parameters
\citep{Dobbs2008,Tasker2009}. These results show that the virial
parameter, $\alpha$ (see Section~2.2) typically lies in
the range of around 0.2 to 10. 

In this paper, we address the question of how molecular clouds can
remain unbound. \citet{Pringle2001} argued that if molecular clouds
are short-lived, with lifetimes comparable to a few tens of Myr, then
they must be formed from a large reservoir of dense interstellar gas,
which may or may not itself be molecular. \citet{DBP2006} has shown that the
formation of the global structure of molecular gas (clouds, spurs
etc.) does not in itself require self-gravity, but that formation can
come about for entirely kinematic reasons. In this paper we take these
ideas a step further and attempt to model the observed properties of
molecular clouds. We self consistently follow the evolution of clouds
in a galactic disc, taking into account cloud collisions and cloud
dispersal by energy input from stellar feedback.
The clouds we consider are of size tens of
parsecs, we are unable to resolve very small clouds. 
The particular properties we try to match are the
observed distribution of the virial parameter $\alpha$, the shapes of
the clouds and their internal structures. We find that
these properties can be matched simply by assuming that those regions
within molecular clouds that become self-gravitating are able to form
stars at some small efficiency (5 -- 10 per cent) which gives rise to
feedback in the form of input of energy and momentum (Section 2).
Thus if say only around 10 per cent of a cloud is bound at any one
time, and those parts form stars at around 10 per cent efficiency, the
problem of the two order of magnitude difference in the star formation
rate identified by \citet{Zuckerman1974} can be overcome (see Section~3). We
demonstrate that with this simple assumption, those structures which
would be identified as molecular clouds are, for the most part, globally unbound, with
properties giving a reasonable match to the data.  

\section{Simulations}
The calculations presented here are 3D SPH simulations using an SPH code
developed by Benz \citep{Benz1990}, Bate \citep{Bate1995} and Price
\citep{PM2007}. 
The code uses a variable smoothing length, such that the density
$\rho$ and smoothing length $h$ are solved iteratively according to
\citet{PM2007}, and the typical number of neighbours for a particle 
is $\sim 60$. Artificial viscosity is included to treat shocks, 
with the standard values $\alpha = 1$ and $\beta = 2$ \citep{Monaghan1997}.
In all the calculations presented here, the gas is assumed to orbit in a fixed galactic
gravitational potential. The potential includes a halo
\citep{Caldwell1981}, disc \citep{Binney1987} and 4 armed spiral component
\citep{Cox2002}. The gas particles are initially set up with a random
distribution, and assigned velocities according to the rotation
curve of the galactic potential with an additional 6 km s$^{-1}$
velocity dispersion. The rotation curve is flat across most of the
disc, with a maximum velocity of $220$ km s$^{-1}$.

We present results from 4 different calculations, as
summarised in Table~1. Run A was already presented
in \citet{Dobbs2008}, and is more simplistic than Runs B, C and D.
The total gas mass is $5 \times 10^9$ M$_{\odot}$ in Run A, and
  $2.5 \times 10^9$ M$_{\odot}$ in Runs B, C and D, and corresponds to
  one or two per cent of the total mass of the galaxy. The surface
  density of the Milky Way is about 12 M$_{\odot}$ pc$^{-2}$
  (it is 10 M$_{\odot}$ pc$^{-2}$ in \citealt{Wolfire2003} excluding helium) thus a little higher
than Runs B, C and D. The mass
  resolution is 1250 M$_{\odot}$ for Run A, and 2500 M$_{\odot}$ for Runs B, C and
D.

For Run A, we allocate particles at
radii between 5 and 10 kpc. The gas is assumed to be a two phase
fluid. The interstellar medium has two
isothermal components, one cool and one warm. We omit thermal
considerations and so there is no transition between the two phases;
the cool gas remains cool and the warm gas remains warm,
throughout. The cool and warm
gas comprise equal mass in the simulations. 
The gas has initial scale heights of 150
and 400 pc in the cold and warm components respectively, but these 
decrease to 20-100 pc and 300 pc with time. The mean smoothing length is 40 pc. 
In Run A, we also include a magnetic field, 
such that the plasma $\beta$ of the cold gas is 4. The
magnetic field is implemented using Euler potentials 
as described in \citet{DP2008}.

In the remaining calculations (B, C and D), we investigate the effect of stellar
feedback. In these cases we allow the ISM to exhibit a multiphase nature
from 20 K to $2\times10^6$ K. The cooling and heating of the ISM is
calculated as described in \citet{DGCK2008}. Apart from feedback from
star formation, heating is mainly due to
background FUV, whilst cooling is due to a variety of processes
including collisional cooling, gas-grain energy transfer and recombination on grain surfaces. 
The gas initially lies
within a radius of 10 kpc, and has an initial temperature of 7000 K. The implementation of
stellar feedback will be described in detail in a forthcoming paper, but a simple
description is included here. The gas is assumed to form stars when a
number of conditions are met, i)
the density is greater than 250 cm$^{-3}$, ii) the gas flow is converging, iii)
the gas is gravitationally bound (within a size of about 20 pc, or 3
smoothing lengths), iv) the sum of the ratio of thermal and rotation
energies to the gravitational energy is less than 1, and v) the total
energy of the particles is negative (see \citealt{Bate1995}). 
If all these conditions are met, we assume that
star formation takes place; there is no probabilistic element in
our calculation. {We do not however include sink particles,
instead we deposit energy in the constituent particles. We present calculations with star formation
efficiencies, $\epsilon$, of 1, 5 and 10 per cent (Runs B, C and D respectively). 
This means that of the mass that satisfies the above criteria, a
fraction $\epsilon$ of the molecular gas contained therein is assumed
to form stars instantaneously and to provide an energy input
(approximately $1/3$ thermal and $2/3$ kinetic energy) of $10^{51}$ ergs per 160
M$_{\odot}$ of stars assumed to form.~\footnote{This corresponds to a Salpeter IMF with limits
  of 0.1 and 100 M$_{\odot}$.} This energy input, combined
  with our cooling and heating prescription, leads to a multiphase
  ISM. In the case of Run C ($\epsilon=5$ per cent), from 150 Myr
  onwards approximately one
  third of the gas lies in the cold, unstable and
  warm regimes. 
  
\begin{table*}
\centering
\begin{tabular}{c|c|c|c|c|c|c|c}
 \hline 
Run & Surface density  & ISM & $\beta$ & $\epsilon$ & No. particles & Time chosen to \\
& (M$_{\odot}$ pc$^{-2}$) & & & per cent & & locate clouds (Myr) &\\
 \hline
A & 20 & Two phase isothermal & 4 & N/A & 4 x 10$^6$ & 130 \\
B & 8 & Multiphase (20 - $2\times10^6$ K) & $\infty$ & 1 & 10$^6$ & 110 \\
C & 8 & Multiphase (20 - $2\times10^6$ K) & $\infty$ & 5 & 10$^6$ & 200 \\
D & 8 & Multiphase (20 - $2 \times 10^6$ K) & $\infty$ & 10 & 10$^6$ & 200 \\
\hline
\end{tabular}
\caption{The different calculations performed are described above. Run
  A was presented in \citet{Dobbs2008}. For
  this two phase isothermal calculation, half the gas lies in the warm
  ($10^4$ K) phase whilst half is cold (100 K). $\epsilon$ is the star
formation efficiency in the calculations with feedback (see text for
details). For Runs C and D, the time we locate the clouds is not
important as they have reached an approximate equilibrium state.}
\label{runs}
\end{table*}

\begin{figure*}
\centerline{
\includegraphics[scale=0.32, bb=0 0 540 450]{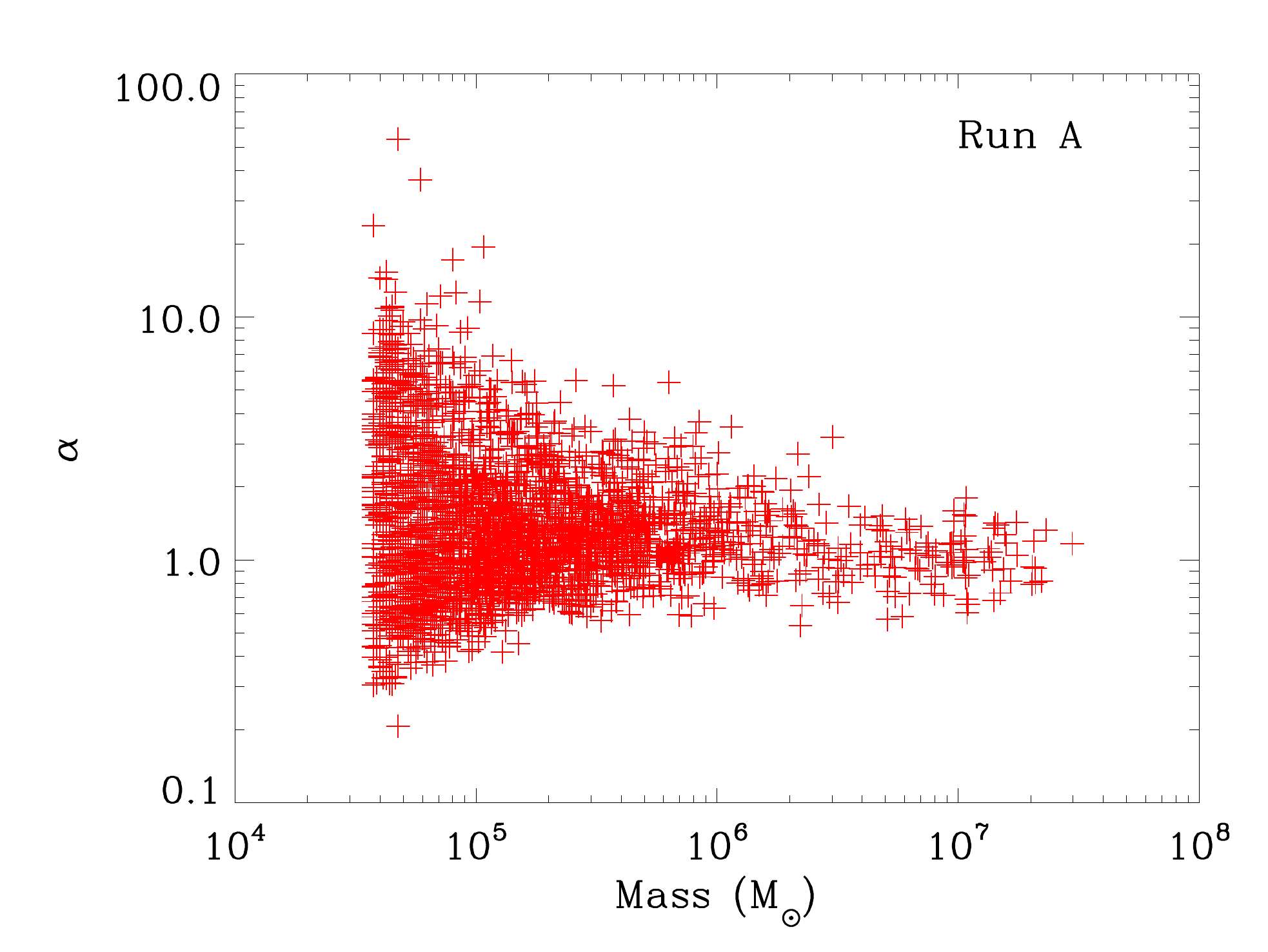}
\includegraphics[scale=0.32, bb=0 0 540 450]{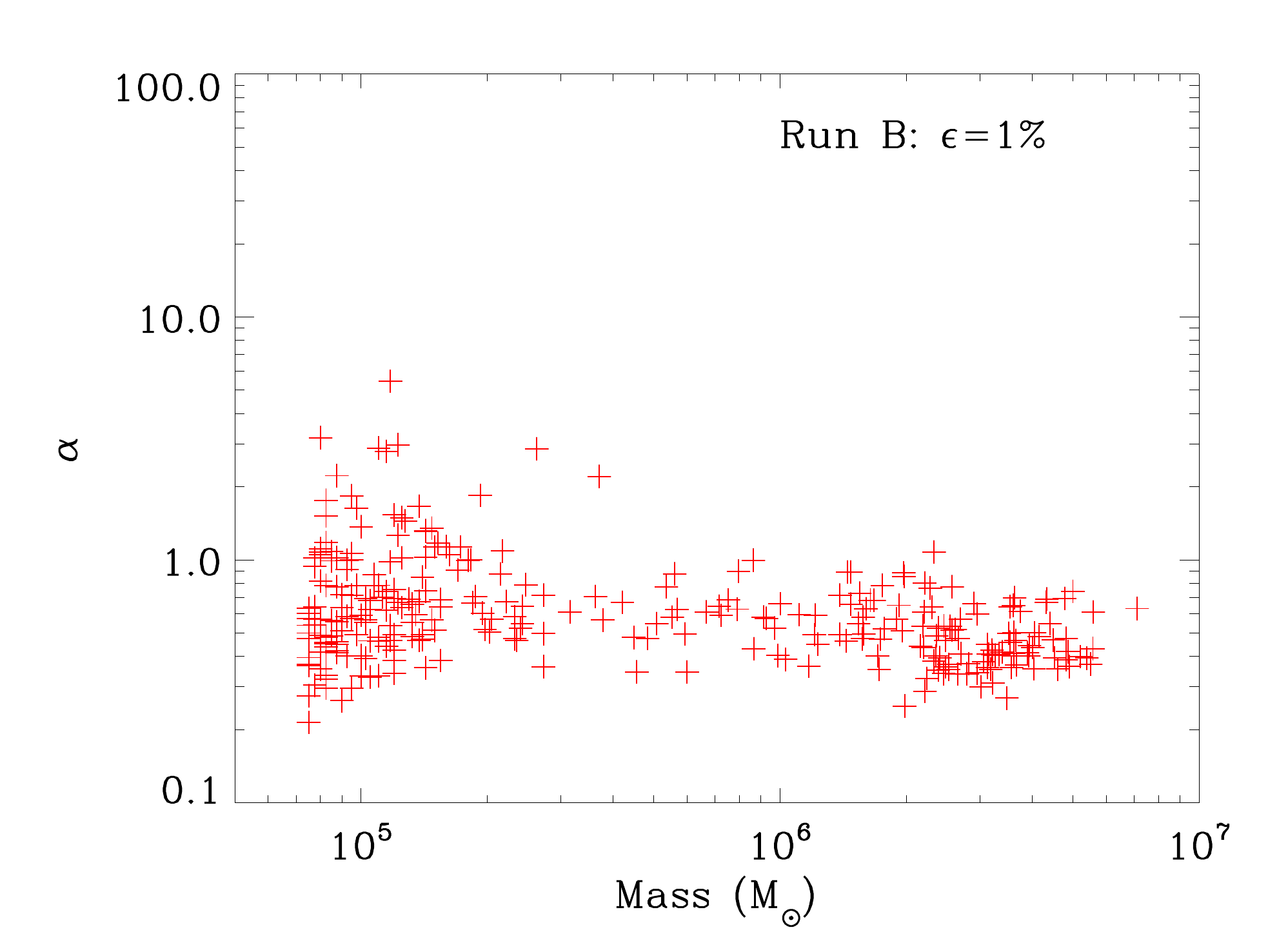}
\includegraphics[scale=0.32, bb=0 0 540 450]{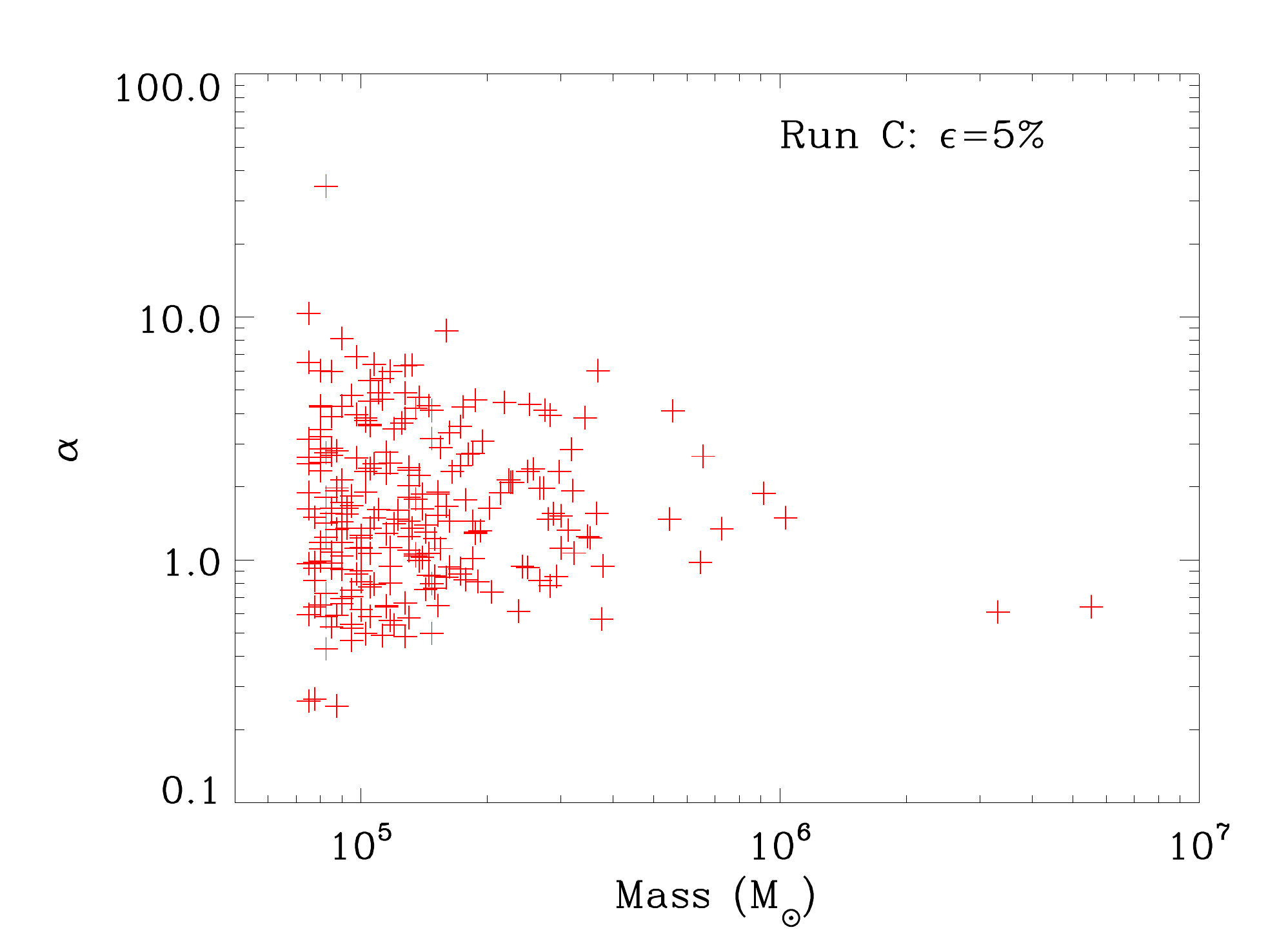}}
\centerline{
\includegraphics[scale=0.32, bb=-60 0 480 400]{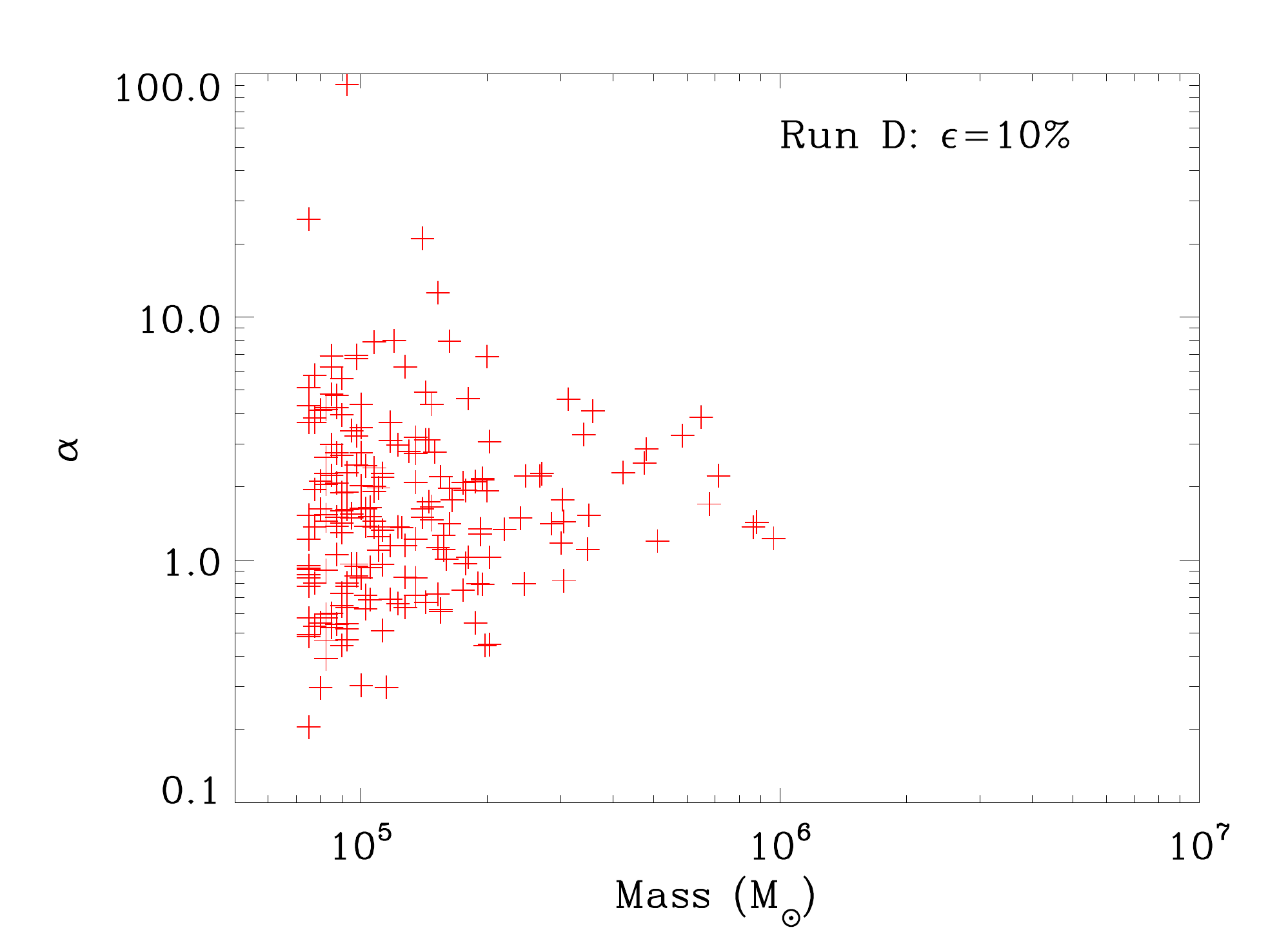}
\includegraphics[scale=0.32, bb=-60 0 480 400]{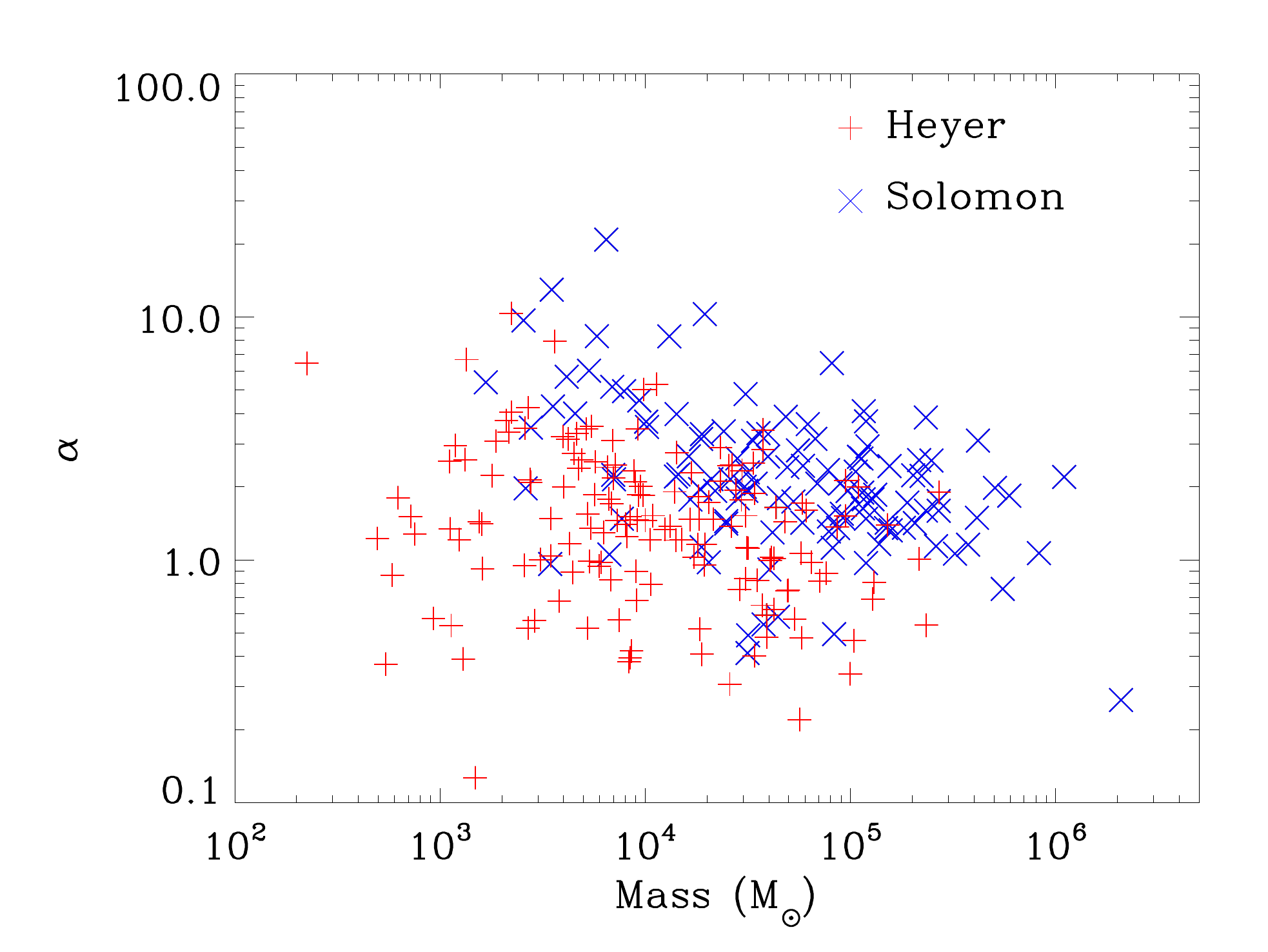}
\includegraphics[scale=0.41, bb=-40 0 500 400]{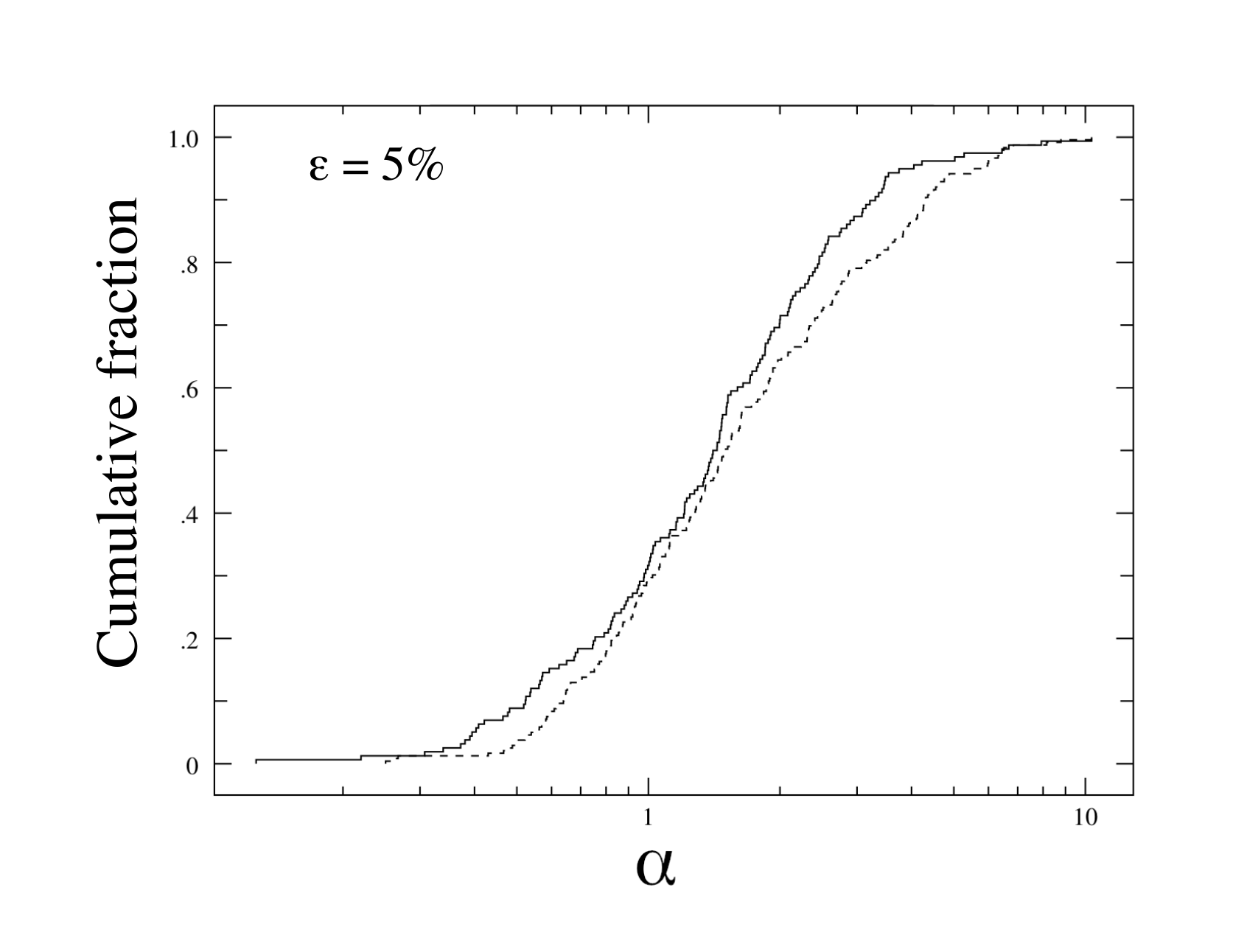}}
\caption{The distribution of the virial parameter ($\alpha$) is plotted
with mass for clouds identified in Run A (top left, with magnetic
fields), in the
calculations with feedback adopting efficiencies of 1, 5 and 10 \%, 
and the \citet{Heyer2009} data (lower middle). We find a
population of predominantly unbound clouds, in rough agreement with the
observations, for the models where localised gravitational collapse is
limited by magnetic fields (Run A), or gravitational collapse occurs
but there is a realistic level of stellar feedback (Run C, top right,
Run D, lower left). There are
many more bound clouds for the case with a very low level of
stellar feedback (Run B, top middle). In the final panel (lower
right), the cumulative fraction of clouds with a given $\alpha$ is
shown for the Heyer data (dotted) and for the clouds from Run C, with
5 per cent efficiency (solid line). The KS test confirms that the distributions of $\alpha$ from the
observations and simulations match, giving $P=0.11$ and $P=0.21$
for Runs C and D respectively.}
\end{figure*}

\subsection{Locating clouds}
We identify clouds using the same method as described in
\citet{Dobbs2008}. We apply a clumpfinding algorithm, which simply
divides the simulation into a grid, and locates cells over a given
surface density. Adjacent cells which exceed this criterion
are grouped together and labelled as a cloud. Clouds which contain
less than 30 particles are discarded, thus clouds in Runs B, C and D
are at least $7.5 \times 10^4$ M$_{\odot}$ (and clouds in Run A $3.75 
\times 10^4$ M$_{\odot}$). The mean number of particles in a cloud is
$\sim 85$ for Runs A, B and C. The properties of the clumps reflect the
total, rather than the molecular gas, but we would typically expect
these clumps to exhibit high molecular fractions. For most of the results we
present, we chose a surface density criterion of 100
M$_{\odot}$pc$^{-2} \sim 4 \times 10^{22}$ cm$^{-3}$. Changing this criterion has little effect on the
distribution of the virial parameter, it merely reduces or increases
the number of clouds selected.  

For Runs C and D ($\epsilon=5$, 10 per cent), we are able to run the simulation for
sufficiently long (300 Myr) that we can calculate cloud properties
when the system has roughly reached equilibrium. However for Runs A
(high surface density, no feedback) and B (1 \% efficiency) we are limited by the
high surface densities reached by a large fraction of the gas.

To illustrate the global structure of the disc in our models, the
column density of the gas in Run C ($\epsilon=5$ per cent) is
shown at a time of 200 Myr in Fig.~2. The dense gas is arranged into
clouds along the spiral arms and spurs extending from the arm to
interarm regions.

\subsection{Virial parameter}
In determining the virial parameters for our clouds, we calculate 
$\alpha$ as shown in Eqn.~1, where $\sigma_v$ is the 
line of sight velocity dispersion and $R$ is defined as the radius of a 
circle with the equivalent area of the cloud. This corresponds to that
used by \citet{Heyer2009}. We take bound clouds as having $\alpha<1$.

In the case with
magnetic fields (Run A), we find largely unbound clouds, 
where local gravitational collapse is prevented by magnetic pressure.
With minimal stellar
feedback (Run B, $\epsilon=1$ per cent), we obtain many more bound
clouds, particularly at higher masses. This clearly disagrees with the observations.
For both Runs C and D ($\epsilon=5$ and 10 \% respectively), we find that the clouds are predominantly
unbound, and the distributions in the virial parameter, $\alpha$ are in
agreement with the observations. This can be seen visually 
and is confirmed by comparing the
distributions of $\alpha$ using the KS test (see also
Fig.~1). Given the uncertainties in determining $\alpha$, 
if we require $\alpha>2$ for an unbound cloud,
 about half of the clouds in Runs C and D
 are unbound. 
There is little change in the fraction of unbound
  clouds with mass (and therefore resolution) in these calculations,
  with the exception of Run D, where feedback is responsible for
  preventing bound, higher mass clouds.
\begin{figure}
\centerline{
\includegraphics[scale=0.3]{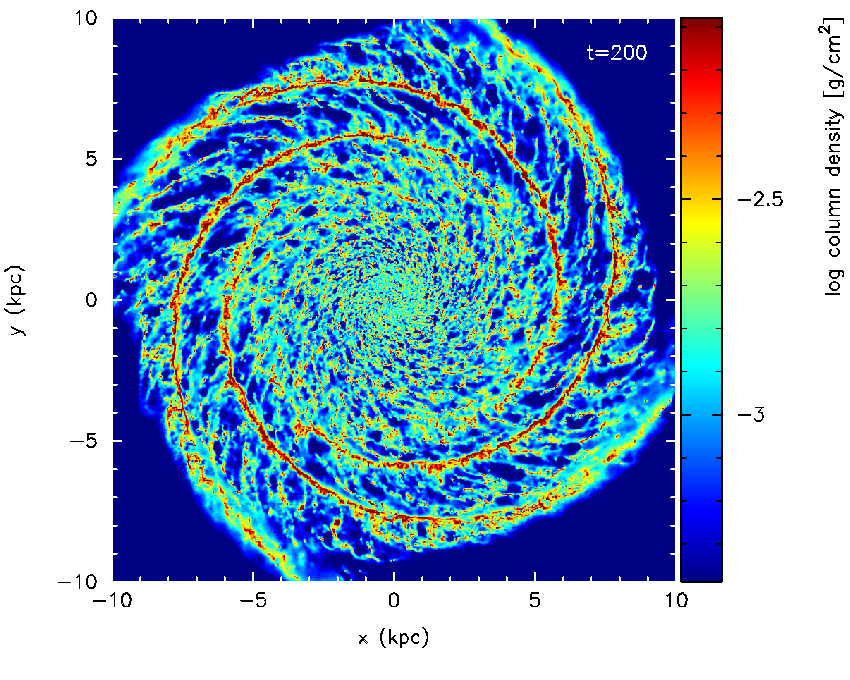}}
\caption{The gas column density is shown for Run C ($\epsilon=5$ per
  cent) at a time of 200 Myr. Dense gas, corresponding to the clouds
  located in the analysis presented here, predominantly lies along the
  arms, and spurs which extend from the arms into interarm
  regions.}
\end{figure}

\begin{figure}
\centerline{
\includegraphics[scale=0.35,bb=100 0 400 400]{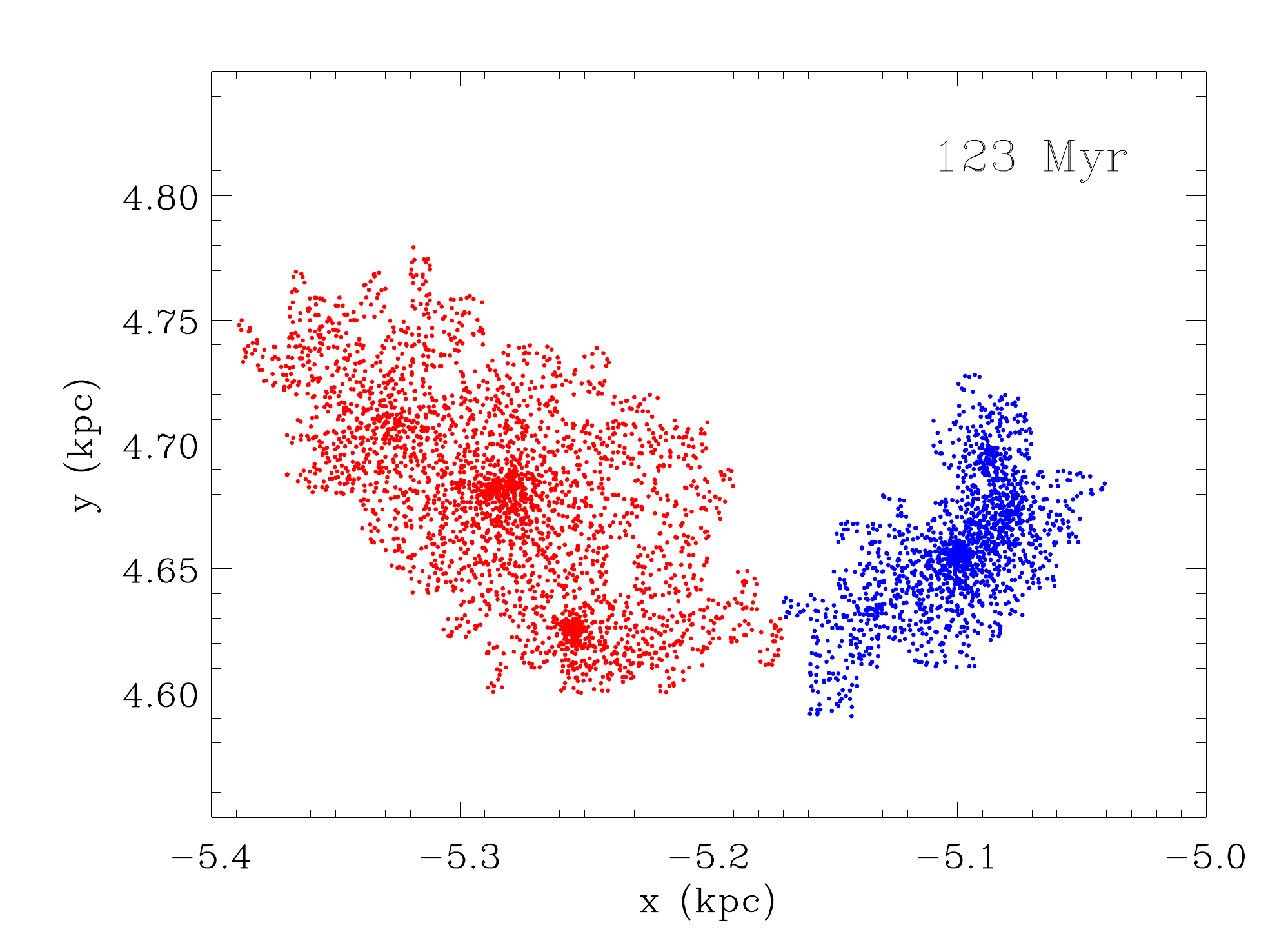}}
\centerline{
\includegraphics[scale=0.35,bb=100 0 400 400]{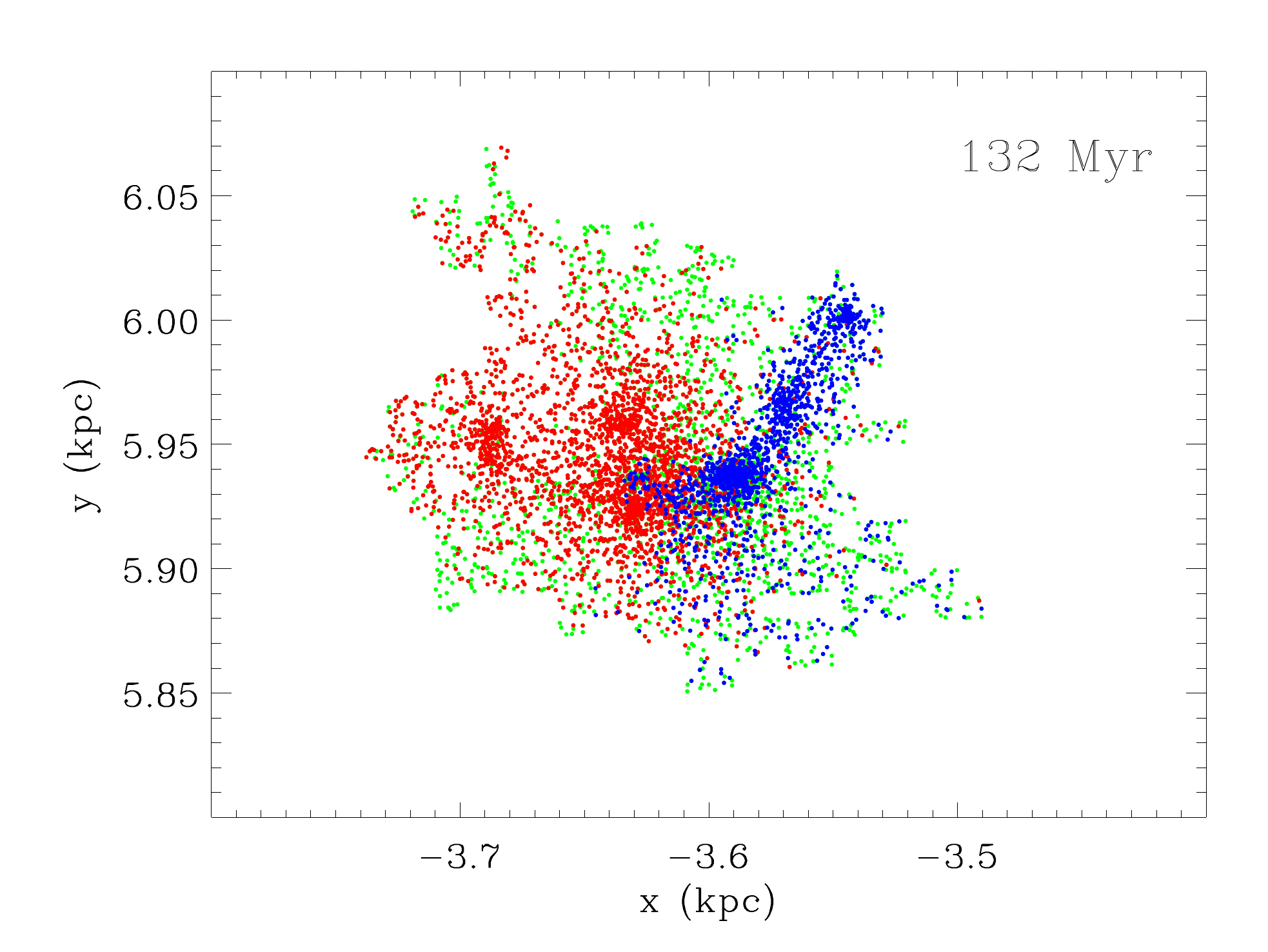}}
\centerline{
\includegraphics[scale=0.35,bb=100 0 400 400]{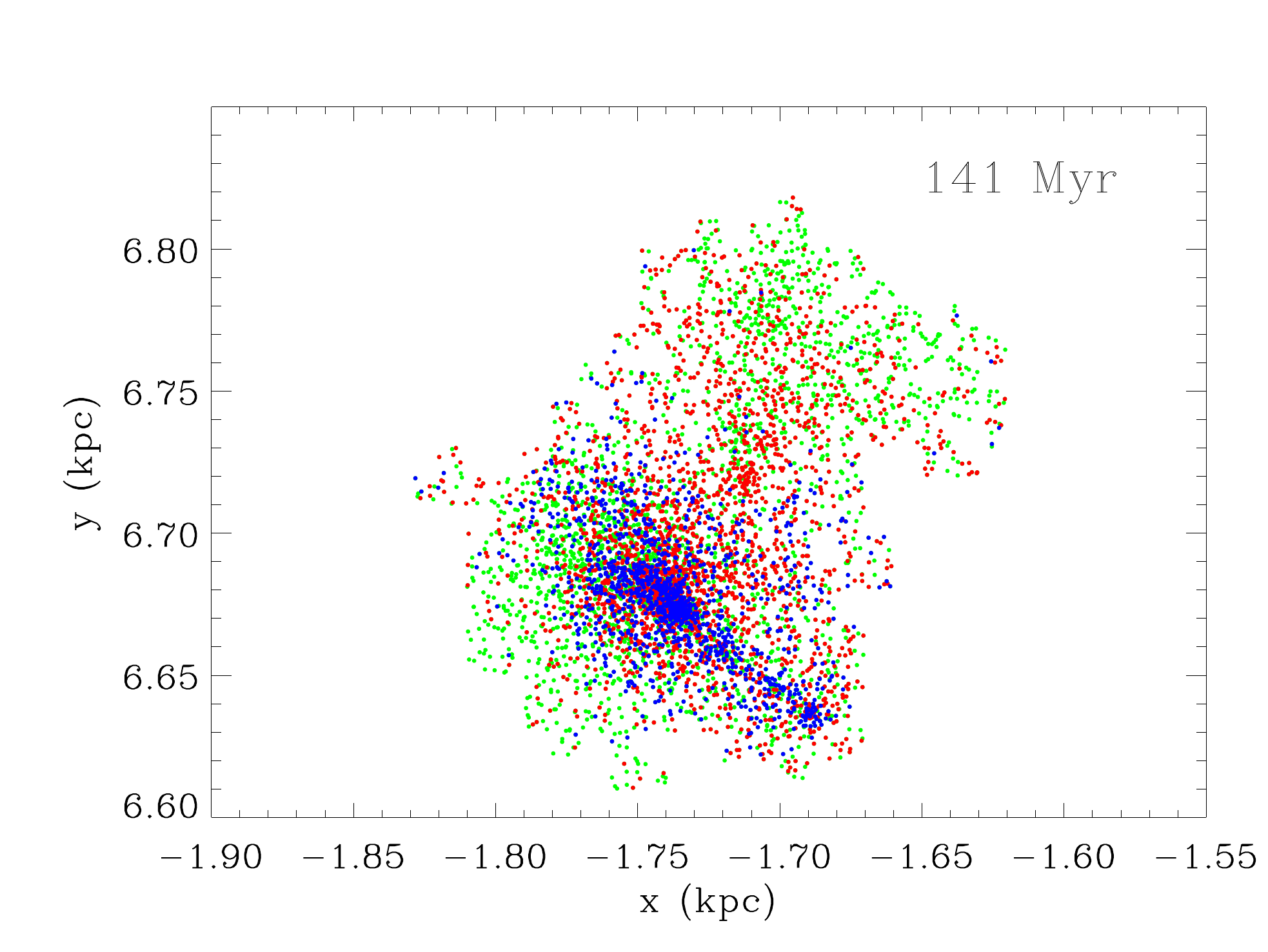}}
\centerline{
\includegraphics[scale=0.37]{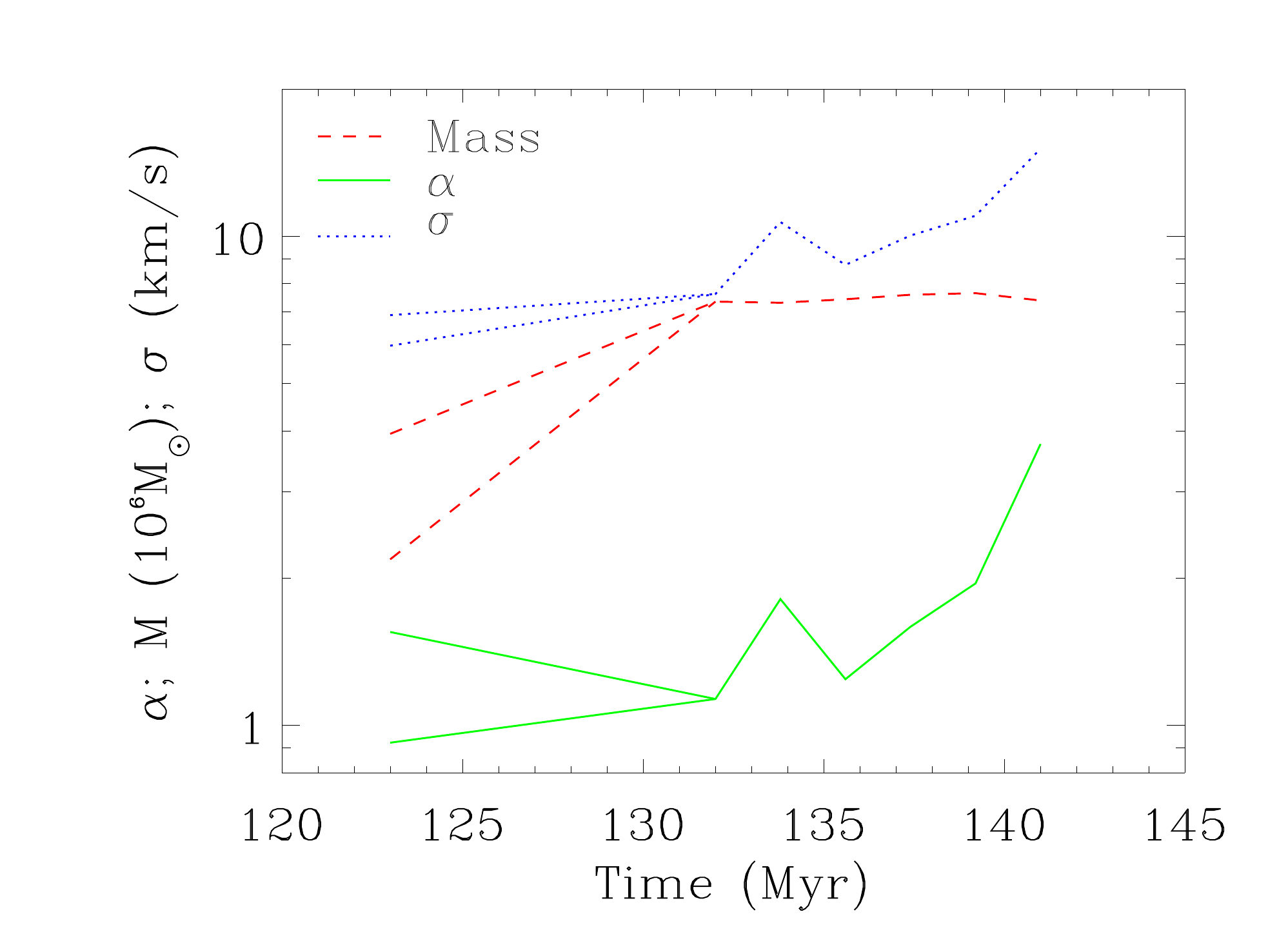}}
\caption{The evolution of a cloud from Run A is shown at times of 123
  (top), 132 (second) and 141 (third panel) Myr. The evolution
  of the mass (in units of $10^6$ M$_{\odot}$), 
  velocity dispersion and virial parameter are shown on
  the lowest panel. The two clouds in the top panel merge,
  and there is a subsequent increase in both the velocity dispersion
  and $\alpha$. They form a single clump of mass $7 \times 10^5$
  M$_{\odot}$. The initial clumps are coloured blue and red, so the
  constituent particles can be traced at later times. The particles
  coloured green are particles present in the clouds at 132 and
141 Myr, which were not present in the clumps at 123 Myr.}
\end{figure}
\subsection{Evolution of individual clouds}
The evolution of an individual cloud is often very complex, involving
collisions, fragmentation and dispersion by feedback. Moreover the gas
which constitutes a cloud can change on relatively short
timescales. Thus studying the evolution of individual clouds, and
establishing why they never, or rarely become gravitationally bound is not
straightforward. 
In this Section we illustrate this behaviour by studying the nature
and development of individual GMCs in the different
calculations.

\subsubsection{Cloud development in Run A}
In Fig.~3, we highlight the contribution of collisions to the internal
velocity dispersions of clouds in Run A. We show a collision between
two clouds in Run A 
(which includes magnetic fields), where small scale
gravitational collapse does not occur. 
After the collision between the
two clouds, the velocity dispersion increases, which means the virial
parameter also increases. The increase in the velocity dispersion is
prolonged because the clouds contain substantial substructure -- the
merging of this substructure is seen in the middle panel. Thus the
collisions between clouds are not really dissipative (as stated in
\citet{DBP2006}); rather the energy is transferred to the internal
motions of the clouds. We also simulated the cloud interaction shown
in Fig.~3 in isolation, and at higher resolution, without
magnetic fields or feedback. This confirmed that the collision induces random
large scale motions which prevent widespread collapse in the cloud
for around 10 Myr, independent of the magnetic field. This
demonstrates that the energy input from cloud-cloud collisions can
be comparable to, or even exceed the energy dissipated, i.e. a collision can lead to an increase in the global
virial parameter. The generation of random velocities is analogous to
that presented in previous work \citep{Bonnell2006,Dobbs2007a}, and
relies on the assumption that the ISM is clumpy on all scales. 
\begin{figure}
\centerline{
\includegraphics[scale=0.3]{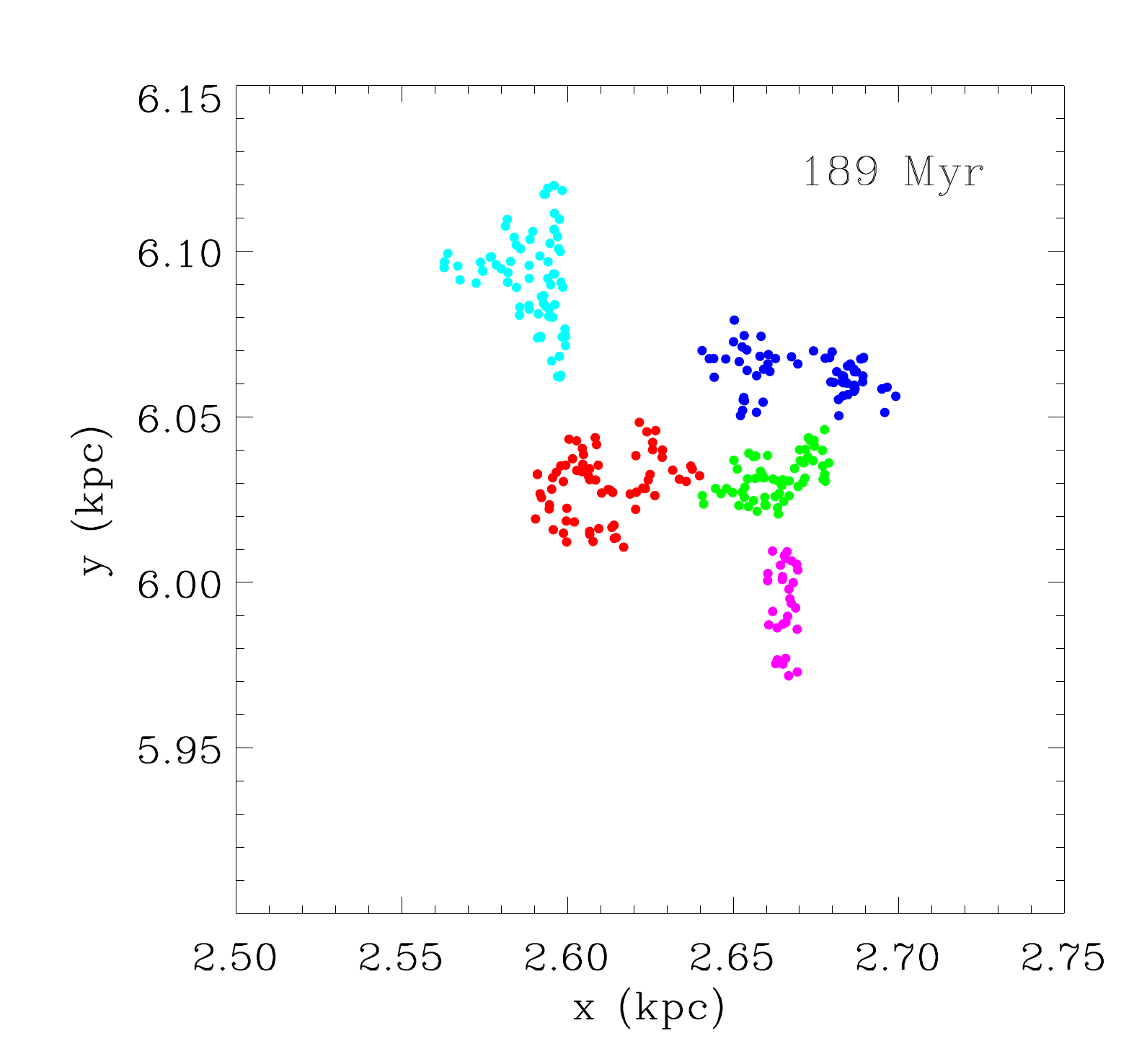}
\includegraphics[scale=0.19, bb=30 30 800 500]{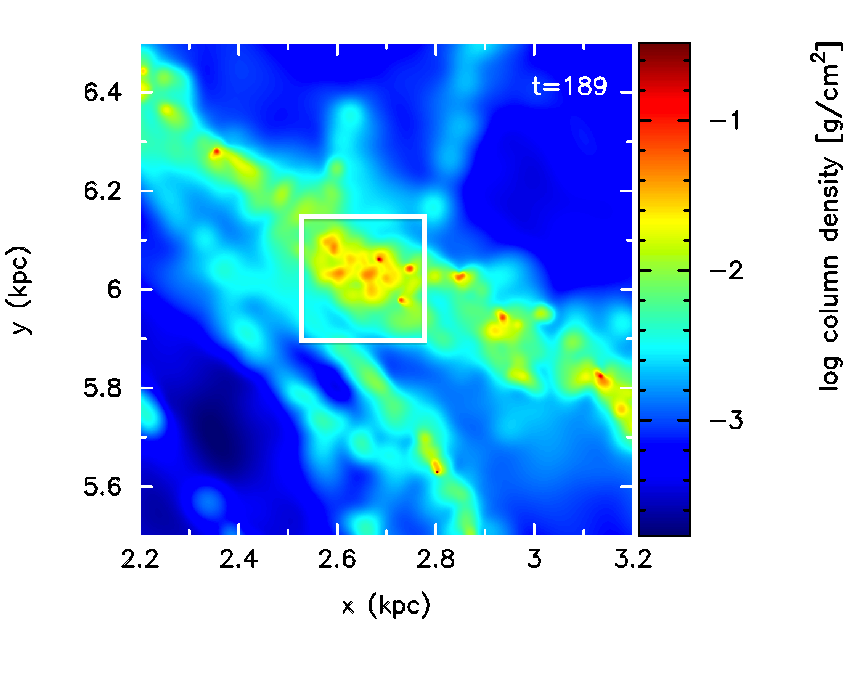}}
\centerline{
\includegraphics[scale=0.3]{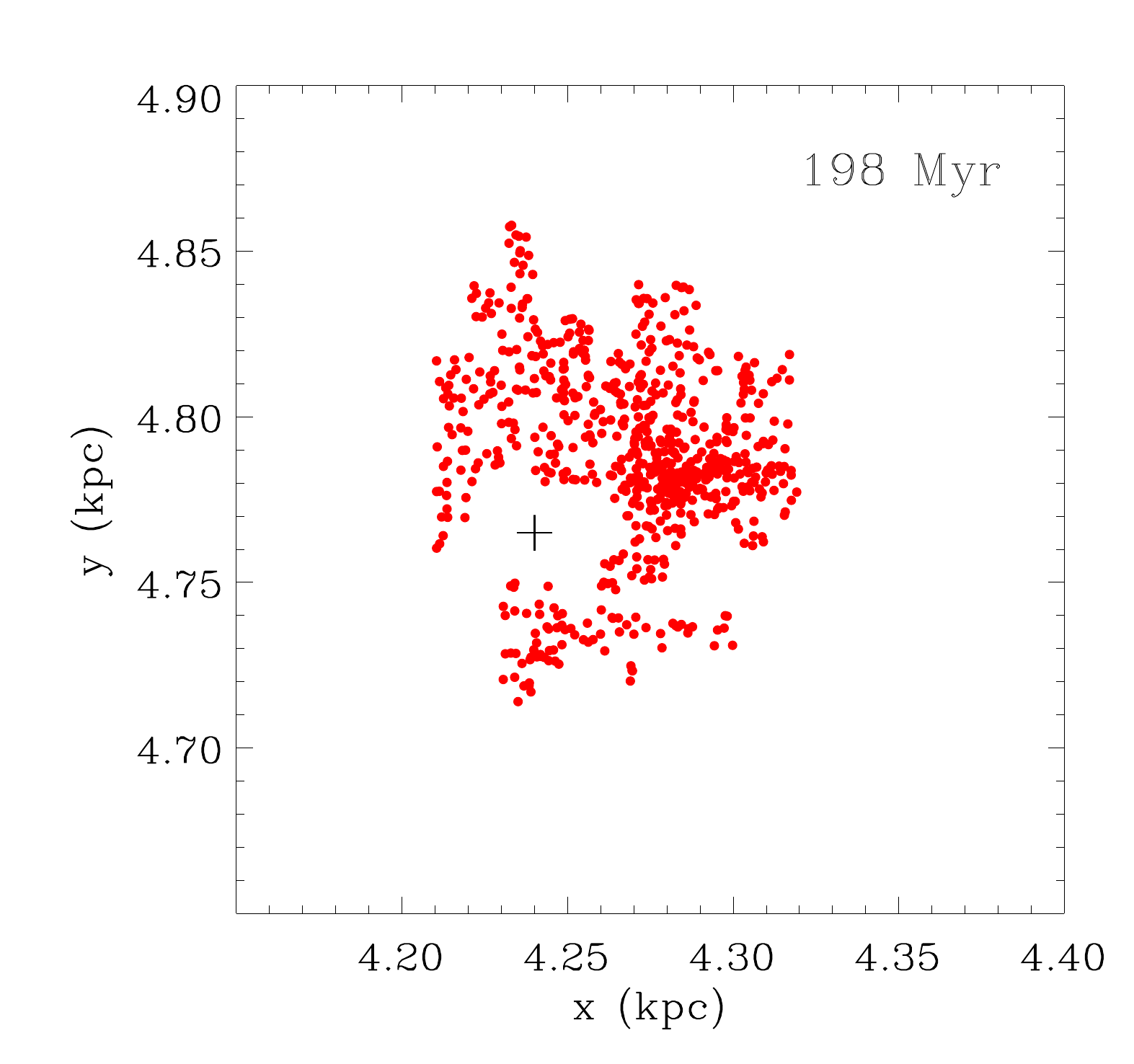}
\includegraphics[scale=0.19, bb=30 40 800 50]{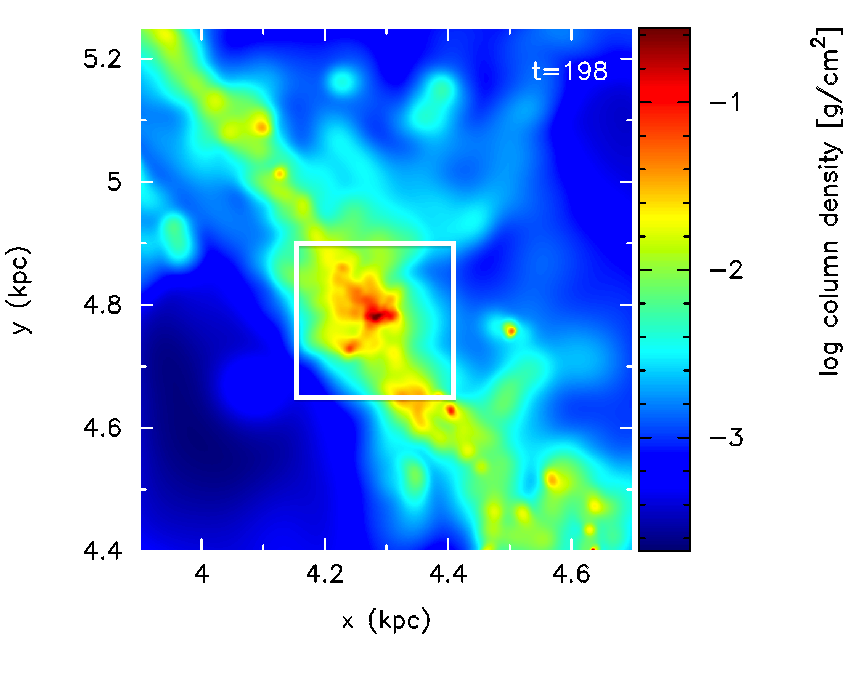}}
\centerline{
\includegraphics[scale=0.3]{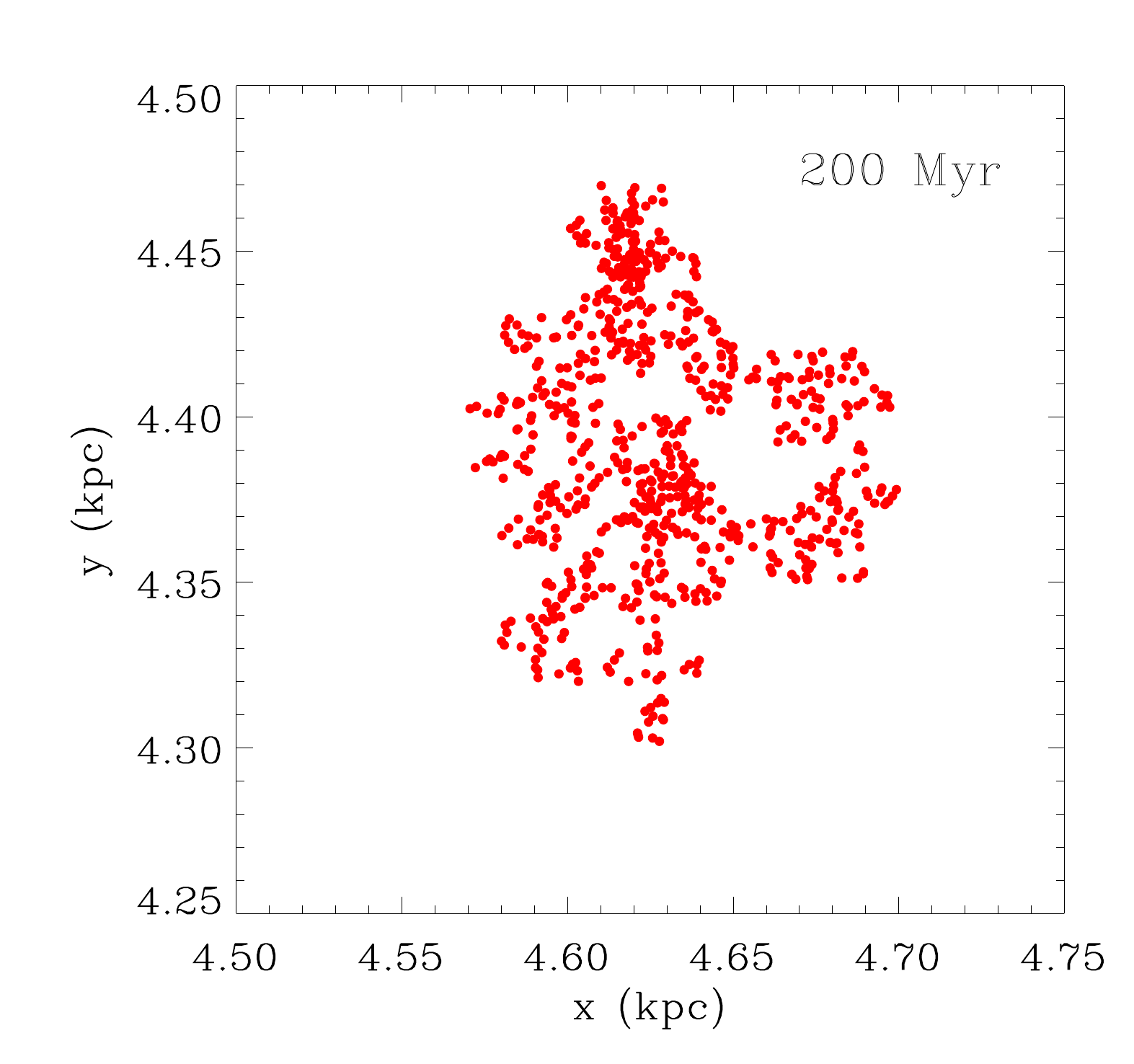}
\includegraphics[scale=0.19, bb=30 40 800 500]{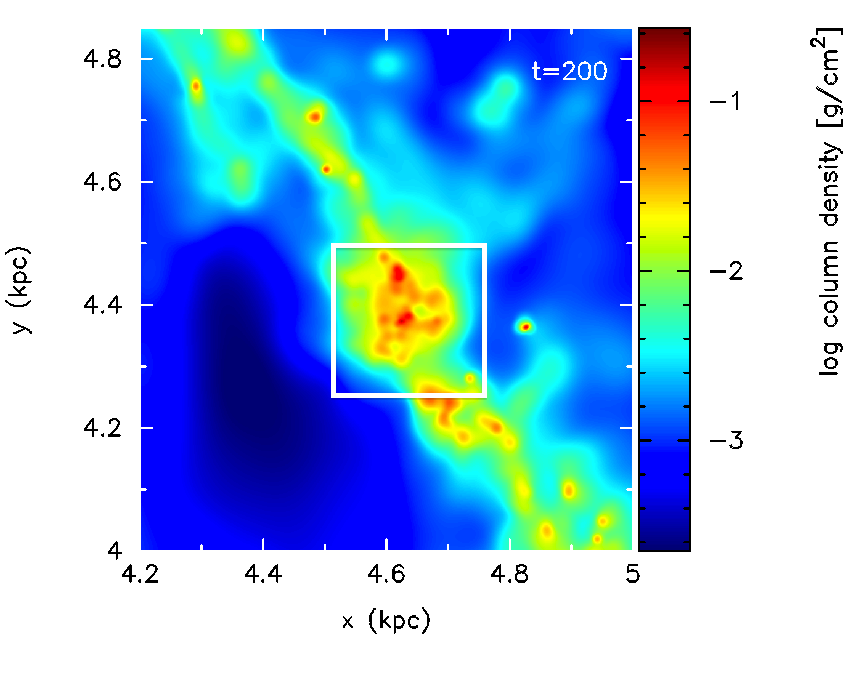}}
\centerline{
\includegraphics[scale=0.3]{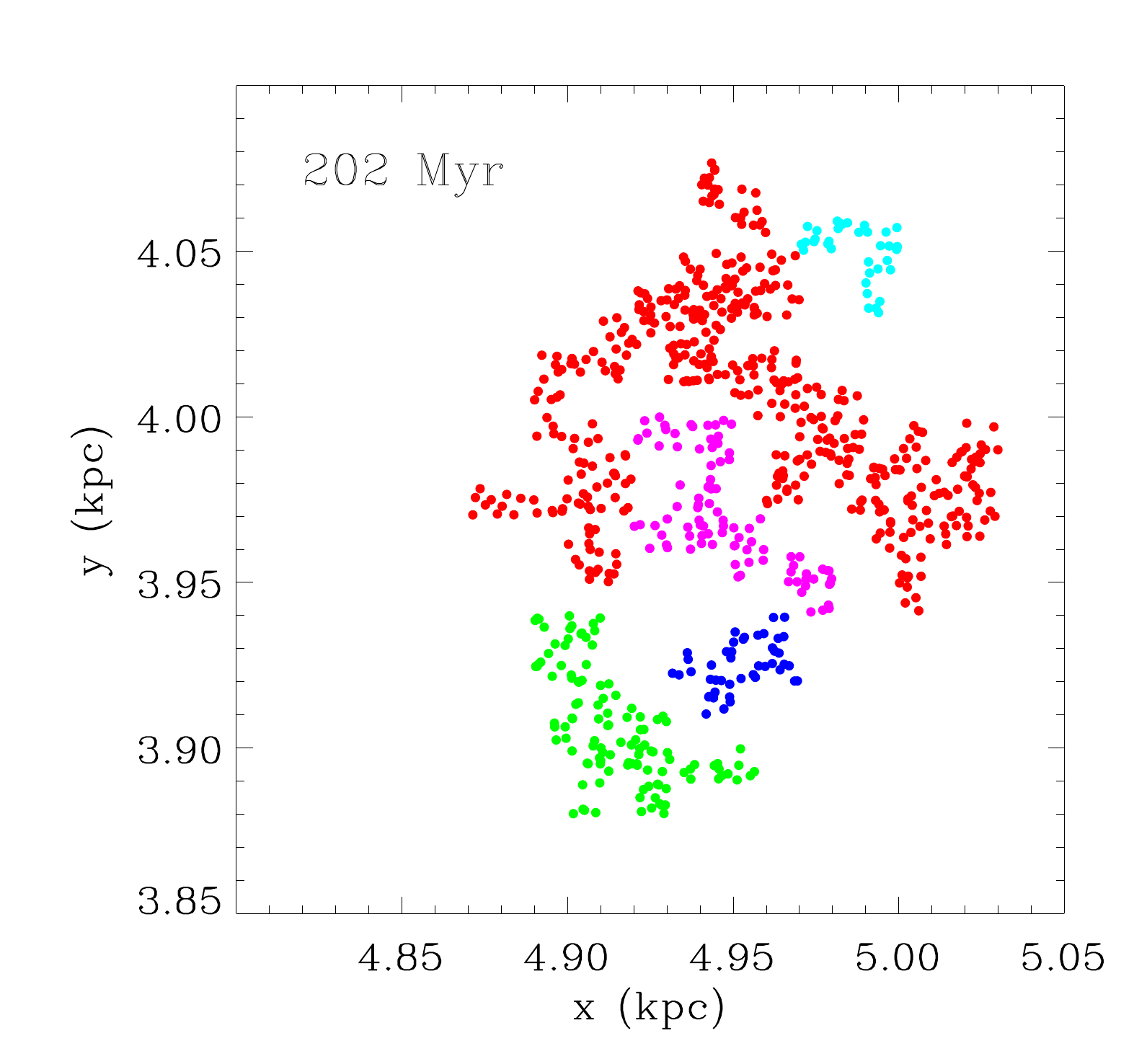}
\includegraphics[scale=0.19, bb=30 40 800 500]{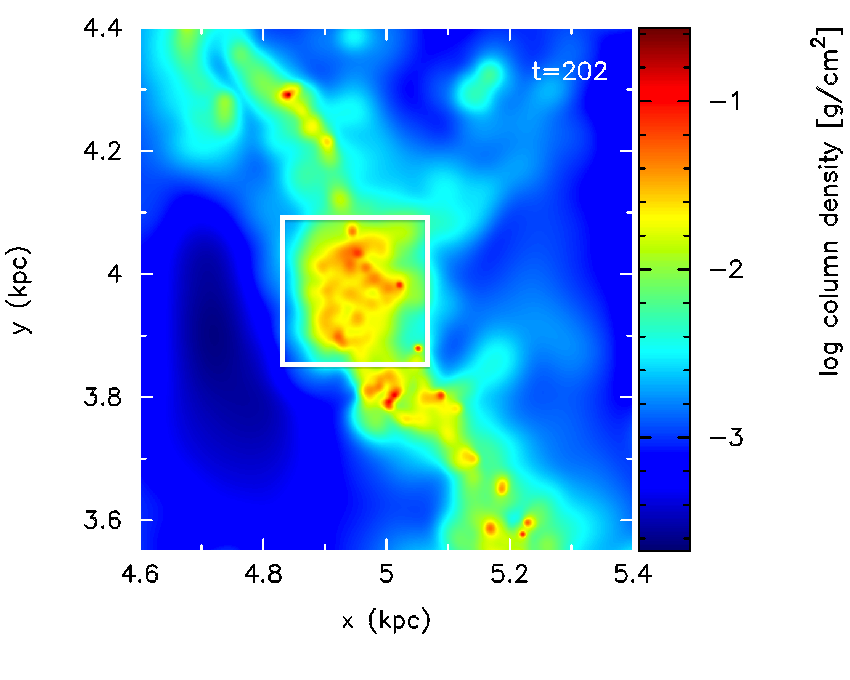}}
\caption{The evolution of a cloud from Run C (with 5 \% efficiency
  stellar feedback) is shown, at 189 (top),
  198 (second), 200 (third) and 202 (fourth panel) Myr.
The cloud is formed by the merger of smaller clumps. Stellar feedback
events (for example the cross in the second panel) then alter the shape of the cloud and finally result in the
separation of the cloud into several separate clumps. 
Separate clumps, (as picked out by the clumpfinding
  algorithm), are shown simply in different colours, but the
constituent particles are not all the same at different times. For
example only 2/3 of the particles in the cloud at 198 Myr are in the
cloud shown at 200 Myr. The
right hand panels show column density images of the clumps and their
surroundings. The white boxes indicate the size of the regions shown on the
left hand panels.  Fig.~5 shows the evolution of $\alpha$, the mass and the
velocity disperions, $\sigma$, 
of the cloud, and the constituent clumps which formed the
cloud.}
\end{figure}

\begin{figure}
\centerline{
\includegraphics[scale=0.45]{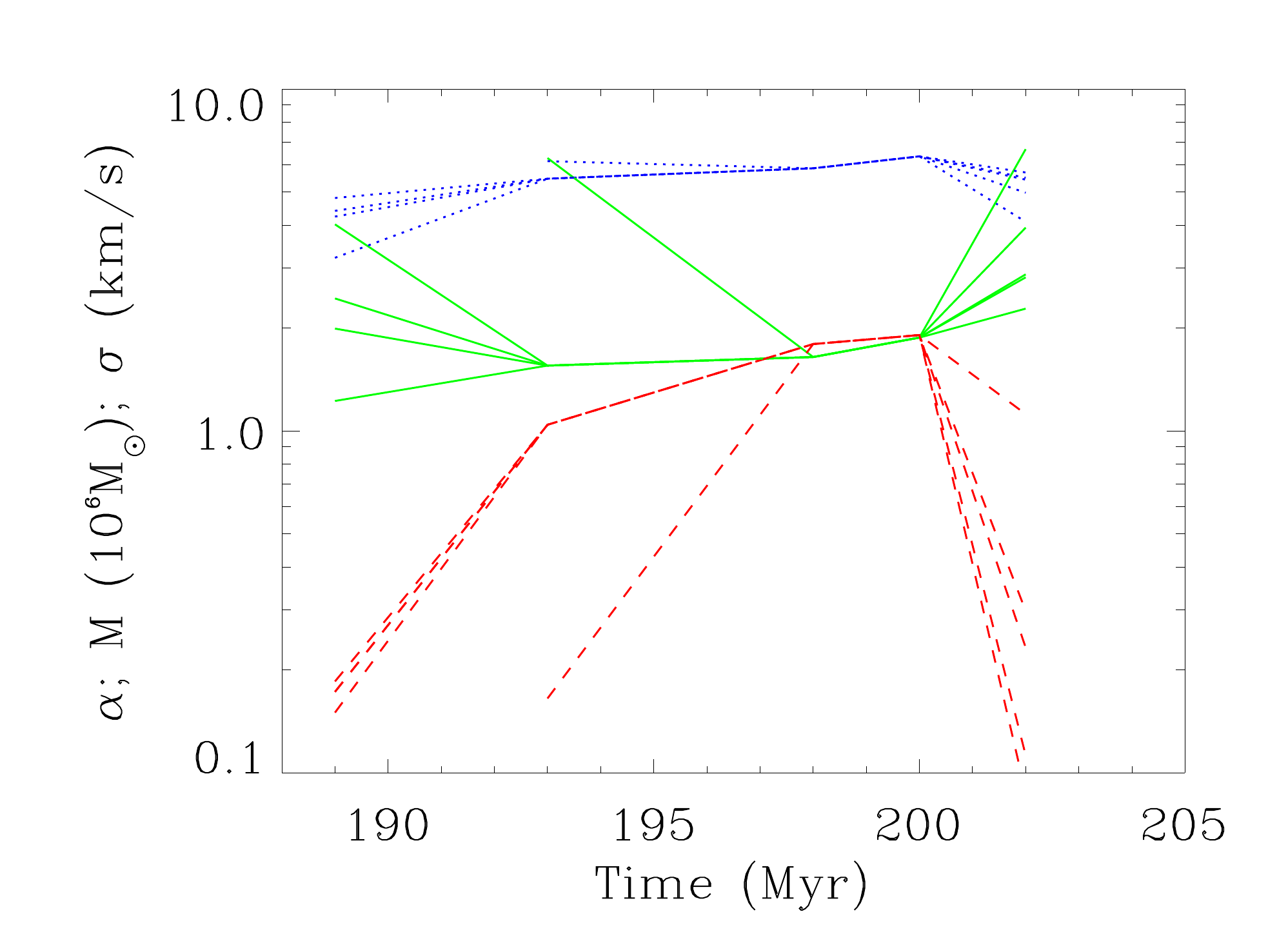}}
\caption{The evolution of $\alpha$ (green, solid), mass (red, dashed
  in units of $10^6$ M$_{\odot}$) and $\sigma$ (blue, dotted) of the clouds
  shown in Fig.~4. Cloud-cloud
interactions have some
role in determining the dynamics of the cloud, and maintaining
$\sigma$, but $\sigma$ is dominated by
stellar feedback. The different lines at 189 Myr (and the line at 193
Myr) correspond to several
clumps merging to form one large cloud (193 Myr to 200 Myr). After 200
Myr, the cloud again splits up into multiple components, due to feedback.}
\end{figure}

\subsubsection{Run C: a multiple cloud interaction with feedback}
In Fig.~4 we show the evolution and interaction of a multiple set of
clouds in Run C ($\epsilon=$5 per cent).
A single cloud was
selected at a time of 198 Myr, and the clouds which contain the same
particles were identified at earlier and later times. We find
that the cloud identified at 198 Myr is formed from the mergers of
several smaller clouds. In the first panel (189 Myr) we can identify 5
separate clouds. As these merge to produce a cloud of $2 \times 10^6$
M$_{\odot}$ some 9 Myr later, the effects of stellar feedback can be seen  
for example in the cloud in the second panel (a bubble blown out by
feedback is indicated by the cross). Feedback plays a large role in
shaping the cloud, and regulating the dynamics. The clouds, as picked
out by the clumpfinding algorithm (left hand plots) show much more
filamentary structures compared to the clouds in Run A
(Fig.~3). By 202 Myr (4th
panel), stellar feedback has succeeded blowing away the top part of
the cloud, and splitting the cloud apart. Over the course of the
plots shown (13 Myr), there are 5 supernovae events in the
cloud. In Fig.~5 we show the corresponding evolution of $\alpha$ and the
velocity dispersion. It can be seen that the velocity dispersion is maintained at about 6
km s$^{-1}$, and the virial parameter, $\alpha$, above unity throughout. Thus
in this case, a multiple cloud collision together with feedback (from
small regions of the cloud which become bound and allow star formation)
maintains the unbound nature of the cloud as a whole.

It can also be seen from Fig.~4 (right hand panels) that the clouds we
identify are part of a larger region of dense gas, which is hundreds
as opposed to tens of parsecs in size. In a galaxy with a high molecular
gas fraction such a feature would correspond to a Giant Molecular
Association (GMA). Whilst these
regions are still unbound in our calculations, they appear to have a longer
lifetime than the GMC sized clouds. 
\begin{figure}
\centerline{
\includegraphics[scale=0.3]{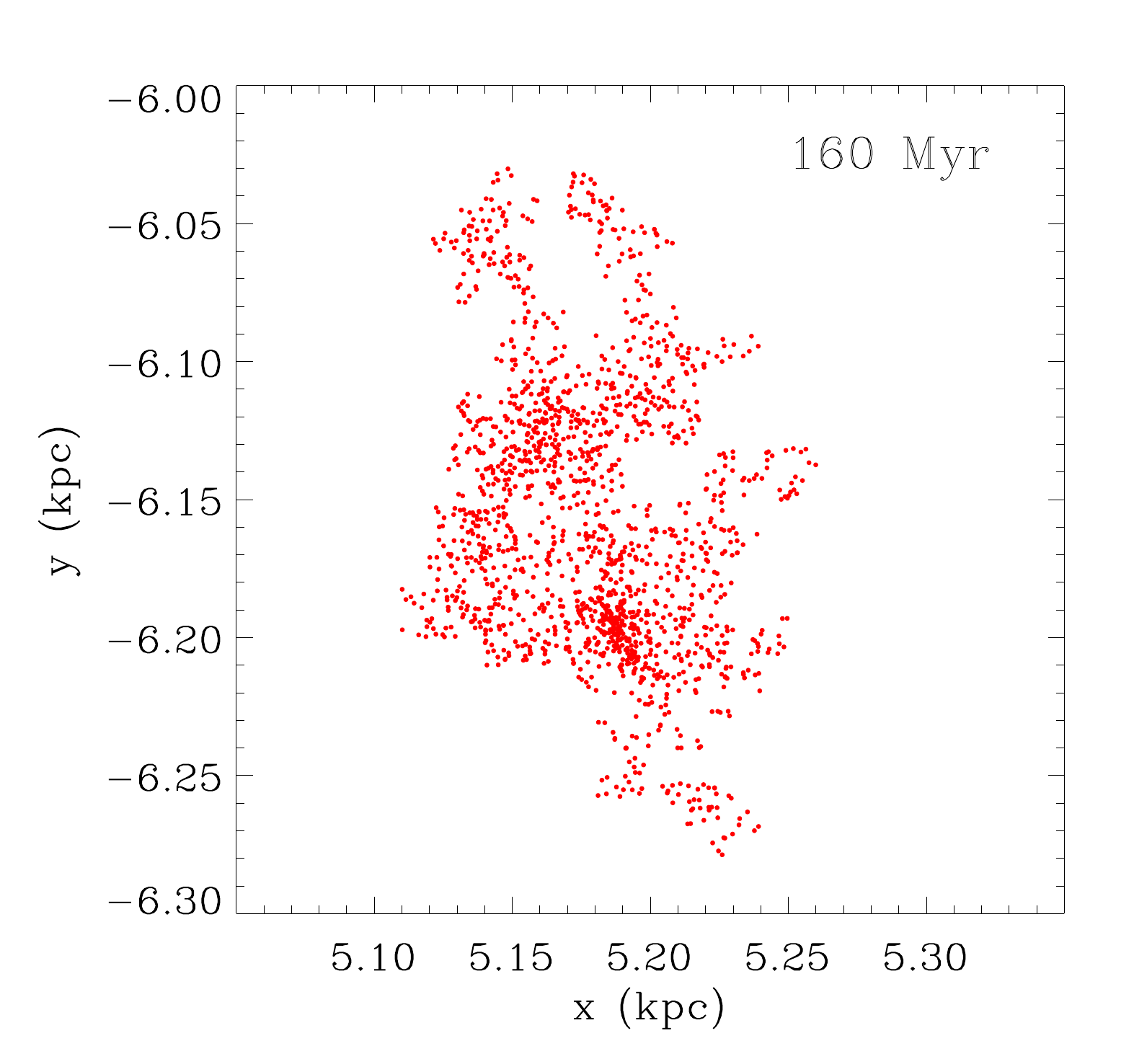}
\includegraphics[scale=0.19, bb=30 40 800 500]{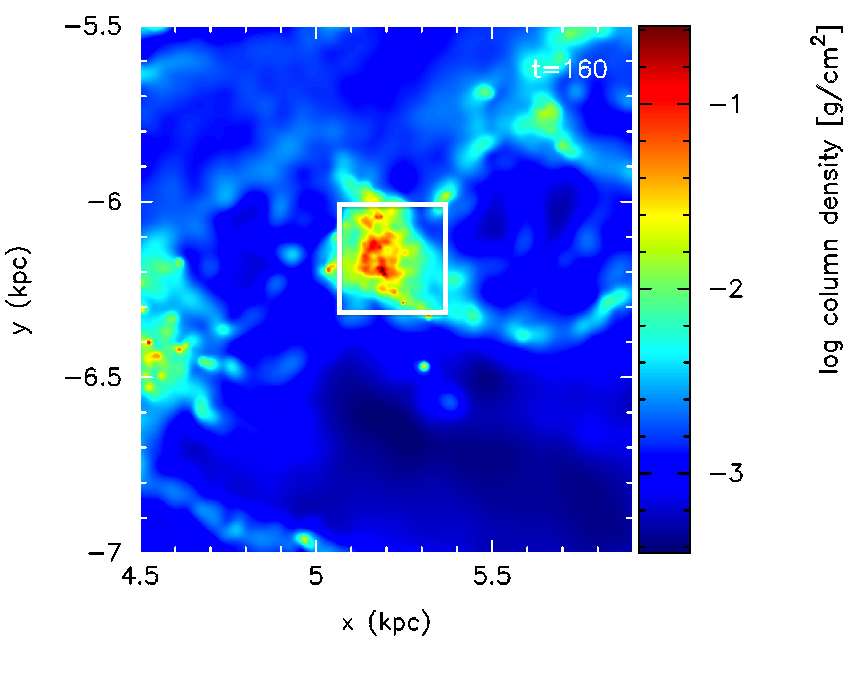}}
\centerline{
\includegraphics[scale=0.3]{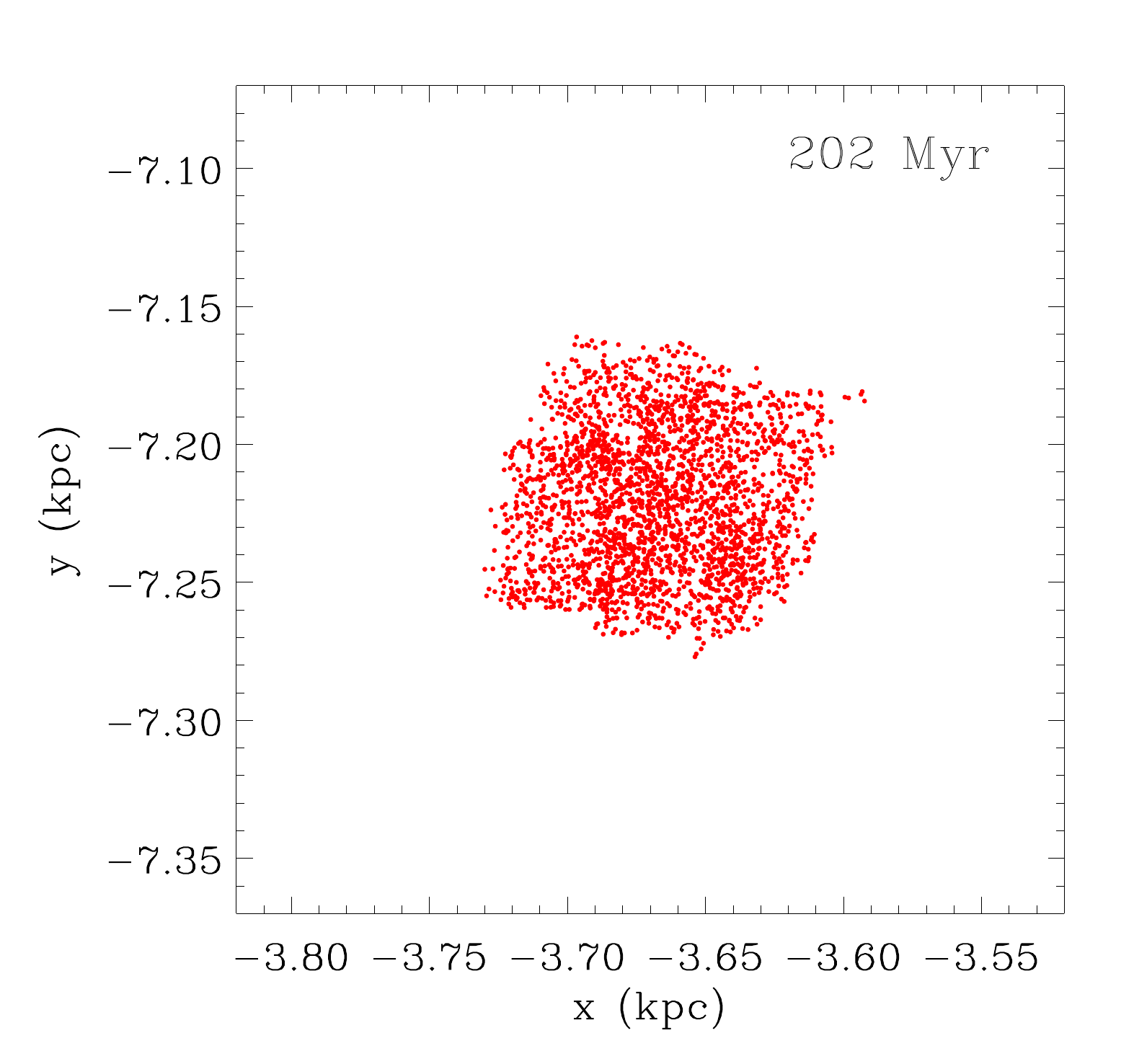}
\includegraphics[scale=0.19, bb=30 40 800 500]{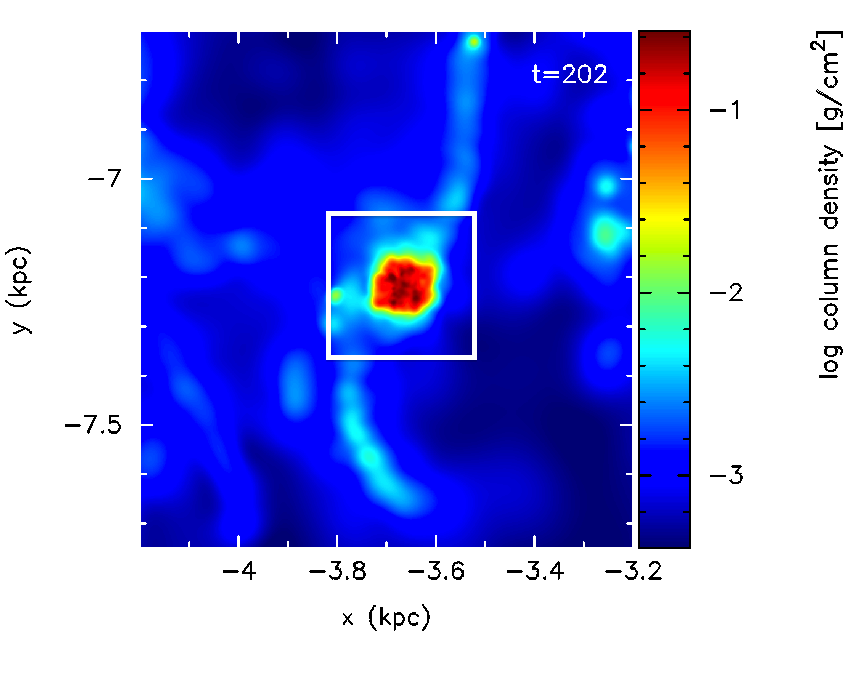}}
\centerline{
\includegraphics[scale=0.35]{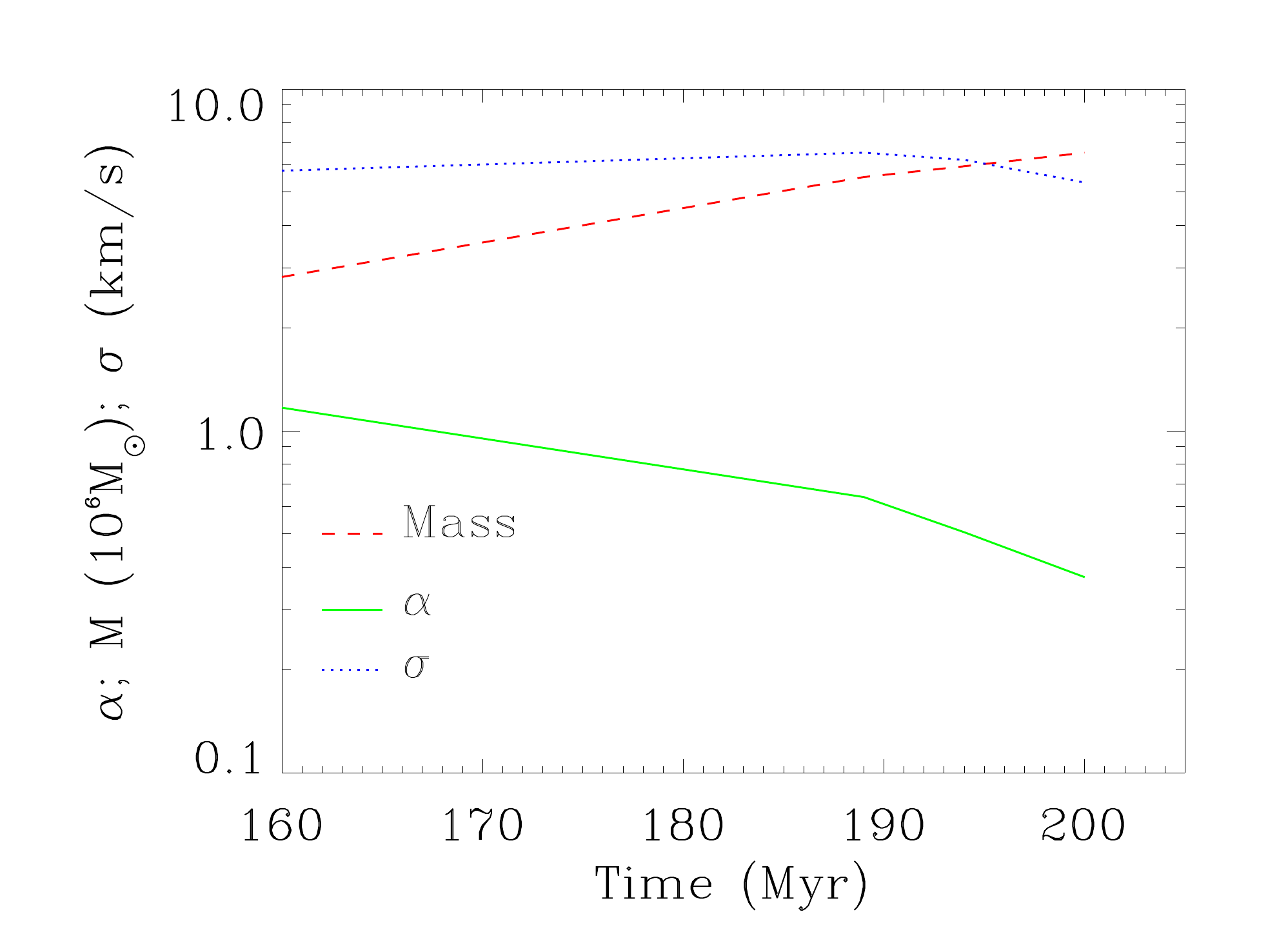}}
\caption{The evolution of a second cloud from Run C, with 5~\%
  efficiency stellar feedback. Unlike the cloud shown in Fig.~4, this cloud is too
  massive to be disrupted by feedback or collisions and becomes
  increasingly more massive, and bound with time. At 160 Myr (top
  panel), the cloud is still filamentary, and marginally unbound. 
By 200 Myr (lower panel), the cloud
contains no filamentary structure and is strongly bound. 
The right hand panels show column density images of the regions
containing the clouds on the left, the white boxes indicating the
sizes of the left hand plots.
The lowest
panel shows the evolution of $\alpha$, mass (in units of $10^6$
M$_{\odot}$) and $\sigma$ for the
cloud.}
\end{figure}

\subsubsection{Run C: the evolution of an isolated massive cloud}
For the calculation, with 5 \% efficiency feedback (Run C), the timescales for the
majority of clouds to merge and become disrupted are relatively short, 
of order several Myr. The exceptions are two longlived bound clouds,
which have masses of $3 \times 10^6$ M$_{\odot}$
and $5 \times 10^6$ M$_{\odot}$ respectively. These are the high mass
points seen in Fig.~1 (top right
plot). We show the evolution of the $5 \times 10^6$ M$_{\odot}$ cloud
in Fig.~6 over a period of 40 Myr. The top
panel shows the cloud at a time of 160 Myr, when the cloud is
clearly irregular in shape. By 200 Myr, the cloud has a much more
regular, quasi-spherical appearance and is not filamentary in any
way. This cloud finds itself in between the spiral arms, and 
does not undergo any significant interactions with other clouds
after 160 Myr. In the lower panel, we plot the evolution of
$\alpha$, the velocity dispersion and the mass. We see that
the cloud is continuing to accrete material, and grow in mass,
becoming steadily more bound. Feedback (with $\epsilon=$5 per cent) 
is insufficient to disrupt
the cloud, although feedback does maintain a constant velocity
dispersion of $\sim$6 km s$^{-1}$. It is likely that this cloud would eventually form a
bound stellar cluster, though we do not attempt to follow this in our
calculation. In Run B ($\epsilon=1$ per cent) there are many more
clouds which display this behaviour.

\subsection{The constituent gas of the clouds}
In the current paradigm of molecular cloud formation and evolution,
GMCs are assumed to be bound objects which consist of essentially the
same gas for the duration of their lifetimes. 
In Fig.~7 we take all the clouds at a given time in Run C
($\epsilon=5$ per cent), and plot
the percentage of gas which remains in a given cloud after 10 Myr.
  
In all cases we find that the constituent
gas of the clouds changes on timescales of $<10$ Myr. We
find that about 50 per cent of clouds are
completely dispersed within 10 Myr. A substantial fraction of clouds are shortlived, either dispersing
to lower densities, merging with other clouds to produce more
massive clouds, or some combination of these processes. There are a
few clouds which substantially retain their identity over a period of
$>$10 Myr.

This highlights that generally the
constituent gas in GMCs is likely to change on timescales of Myr. This
may mean that discussing clouds lifetimes, which are thought to be 20-30 Myr
\citep{Leisawitz1989,Kawamura2009}, may not make
sense. A cloud seen after 30 Myr may not be a counterpart to
any cloud present at the current time. In our calculations this is
only true for the most massive clouds. 

Thus we see that the clouds tend
to display a variety of behaviours. The relatively low mass GMCs undergo frequent
collisions, are readily disrupted, and $\alpha$ will change
accordingly. Even if the cloud becomes bound, it may undergo another
dynamical interaction on a short timescale, and become unbound. In
contrast the more massive clouds undergo less dramatic behaviour.
There are relatively few clouds of this mass, so they
very rarely undergo collisions with objects of a similar size, whilst
above a certain mass they are not so easily torn apart by feedback.

\begin{figure}
\centerline{
\includegraphics[scale=0.35]{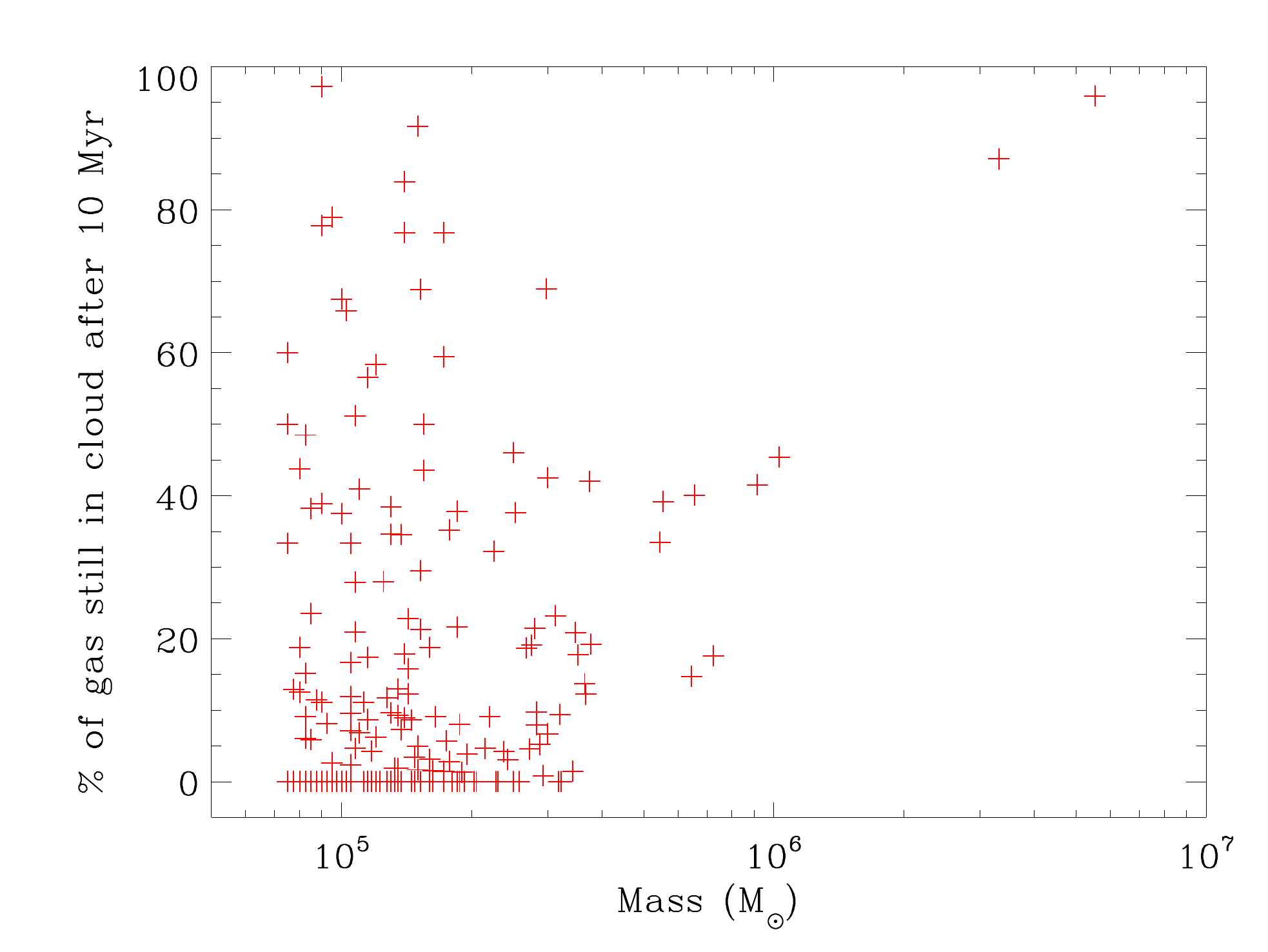}}
\caption{The percentage of gas which still lies in the same cloud after
  10 Myr is plotted versus the cloud's mass for clouds identified in Run C (with stellar
  feedback and $\epsilon=5$ per cent). This percentage is
  calculated by locating clouds at 2 timeframes, 10 Myr apart. We
  search for the constituent gas particles of a given cloud in the clouds
  present 10 Myr later. Sometimes there may be more than one cloud
  containing the particles of an earlier cloud (and in some instances
  no clouds!), in which case we
  select the cloud which has the maximum of the particles the same
  as the cloud at the earlier time frame. 
  In 35 per cent of cases the clouds are completely
    disrupted, whilst the median amount of gas remaining in the cloud
    is 22 per cent.}
\end{figure}

\subsection{The shapes of  clouds}
From our calculations with different levels of feedback, we
obtain distributions of clouds which are predominantly unbound (Runs C
and D, with 5 and 10 \% efficiency) or bound (Run B, with 1 \% efficiency).
We have demonstrated that the observations are most likely fit by a
distribution of mainly unbound clouds.

In addition, we find that, the bound clouds
are much more regular, spherically shaped, whilst the unbound
clouds have very irregular shapes. \citet{Koda2006} carried out a
study of Galactic molecular clouds and determined the aspect ratios of
the clouds. Their results are shown in Fig.~8, and indicate the most
common aspect ratios are between 1.5 and 2. We also show in Fig.~8 the
distributions of aspect ratios for the calculations with low feedback
Run B (predominantly bound clouds), and the higher feedback case, Run
C (predominantly unbound clouds).

In Run C (centre panel), with feedback efficiency of 5 per cent, the distribution of
aspect ratios has reached an equilibrium state, and is found to be
similar to observations, with a peak at about 1.5. 
In Run B (left panel),  with 1 \% efficiency, equilibrium has yet to be achieved, and more and more
clouds have aspect ratios of around unity with increasing time. As
we would expect, the virialised clouds tend to have aspect ratios close to
unity. The prominent peak at aspect ratios of unity does not agree
with the observations, reconfirming our previous conclusions that the
simulations with efficiencies of 5 -- 10 per cent are in best
agreement with the observations.
\begin{figure*}
\centerline{
\includegraphics[scale=0.32, bb=0 0 500 450]{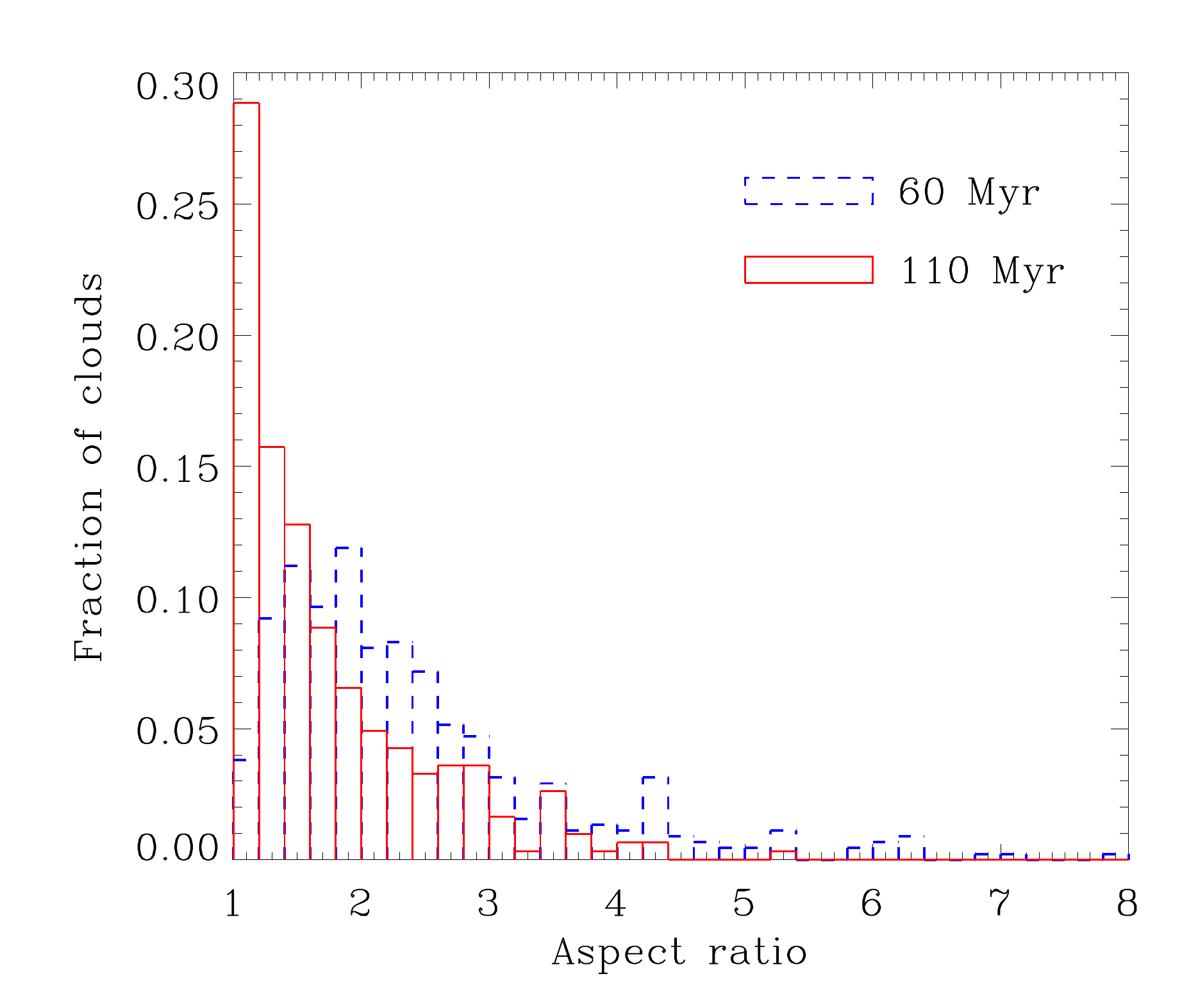}
\includegraphics[scale=0.32, bb=0 0 500 450]{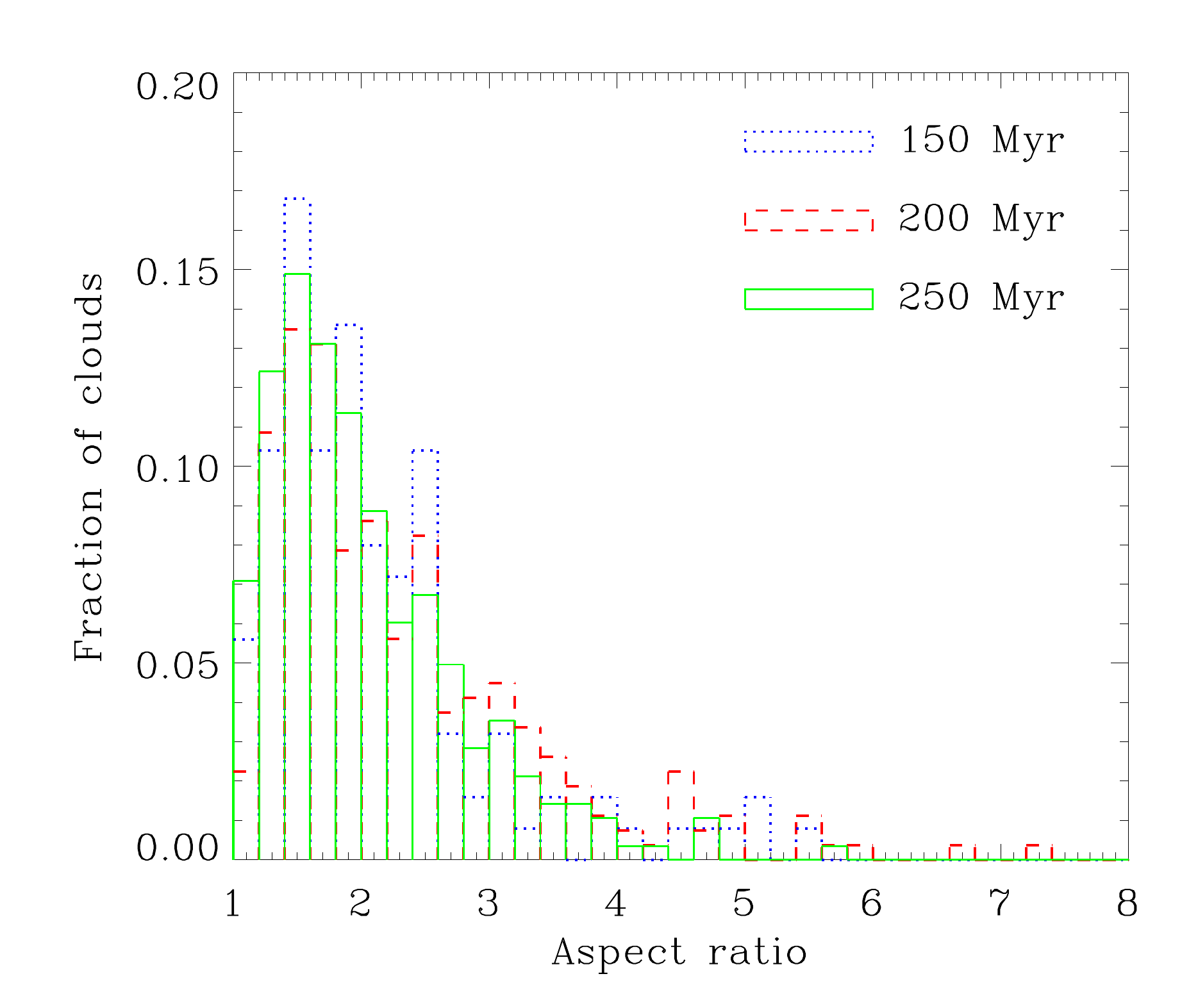}
\includegraphics[scale=0.32, bb=0 0 500 450]{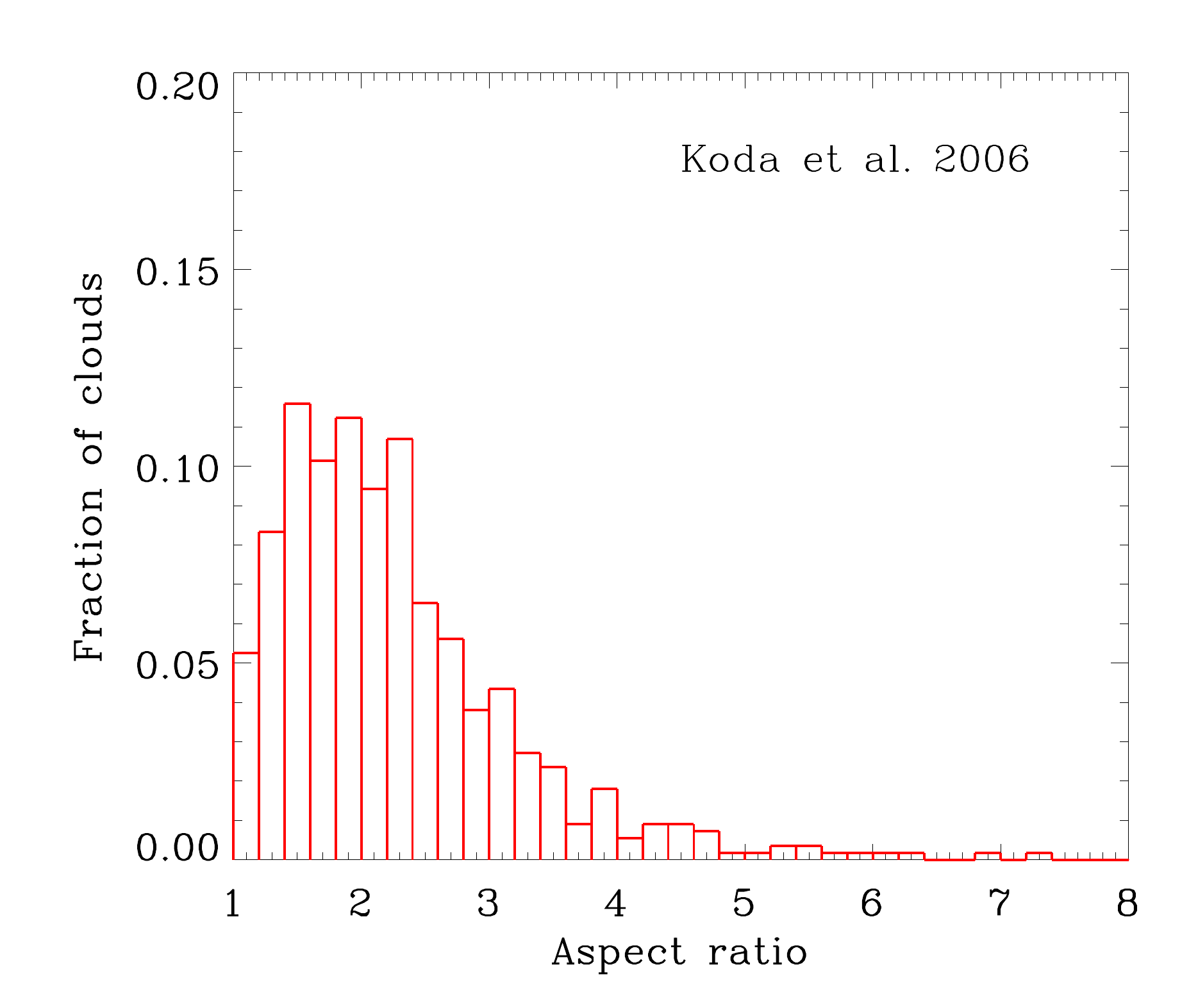}}
\caption{The distribution of aspects ratios of the clouds is shown for
  the models with 1 \% efficiency feedback (Run B, left), and 5 \% efficiency
  feedback (Run C, centre). The distribution of aspect ratios for Galactic
  clouds is shown on the right \citep{Koda2006}. The clouds for the 5
  \% efficiency case (centre) reasonably match the observations,although even
  in this case our clouds are slightly more peaked towards low aspect
  ratios than the observations. The
  distribution does not change with time, once equilibrium has been
  established. With 1 \% efficiency (left), the
  distribution evolves to a strong peak at 1, in definite contradiction to the observations.}
\end{figure*}

\section{Discussion}
In this paper we have addressed the recent observational
evidence that most molecular clouds within the Galaxy are not
gravitationally bound. This evidence contrasts with the original claims of
\citet{Solomon1987}, long since propagated into the field, that
molecular clouds are bound and in virial equilibrium. The idea that
molecular clouds are bound entities has also been taken as a
starting point for many theories of star formation. 

Of course, for star formation to take place it is necessary that {\it
some parts} of a cloud be self-gravitating and able to undergo
collapse.  
What the observations seem to indicate however is that it is not
necessary for the cloud {\it as a whole} to be dominated by
gravity. With this in mind, we present here simulations of the ISM in
a galaxy with a fixed spiral potential. By including simple heating
and cooling of the gas, we are able to identify those parts of the ISM
which are dense enough to represent molecular gas, and so are able to
identify what would be observed as molecular clouds. We allow the
parts of those clouds which are sufficiently dense and sufficiently
gravitationally bound to notionally form stars.
Because such clouds are generally highly
inhomogeneous entities, within them there will be some regions (in our
simulations typically representing only $<30$ percent of the mass) which are
gravitationally bound, and within which star formation takes place. 
  This star formation is
taken to manifest itself as an input of energy and momentum into the
surrounding gas. The global galactic star formation rate is in accordance
with the results of \citet{Kennicutt1998}.

Using this simple, and highly idealised, input physics we are able to
reproduce both the observed distribution of virial parameters of
molecular clouds in the Galaxy (with most of the clouds being unbound)
and also the observed distribution of cloud shapes (in terms of their
aspect ratios). We find that the velocity dispersions within clouds
are maintained not just by feedback from star formation but also by
collisions between non-homogeneous clouds (cf. \citealt{Dobbs2007a}).
However with no, or little feedback, the clouds are predominantly bound and
quasi-spherical (as found in Run B and by \citealt{Tasker2009}), in disagreement
with observations.

We
also find that the constituent material of a typical cloud only remains
within that cloud for a timescale of around 10 Myr. Thus for timescale
much longer than this, the concept of a cloud lifetime is no longer
meaningful. 

 We note that the properties and lifetimes of clouds 
  depend somewhat on the size scales considered. Above some surface
  density threshold, we would expect to start selecting bound regions
  within a GMC, and therefore we would obtain a higher fraction of bound objects.  
 We have not considered the properties of larger
  GMAs either. The fraction of unbound clouds also depends on how we
  define $\alpha$, and what threshold we use. However we note that for
  our clouds, even if $\alpha$ is low, external energy input from
  collisions, and feedback sources within a cloud can act to increase
  $\sigma_v$, and therefore $\alpha$. Thus the main difference between
  our models and previous analaysis, for example that presented by \citet{Ball2010}, is that
  collisions and feedback play a much more important role. 

In summary, the idea that all molecular clouds are gravitationally
bound entities is neither observationally viable, nor theoretically
necessary. It is no real surprise that most molecular clouds
identified in the Galaxy are globally unbound, and that the rest are at most
only marginally bound. 

\section{Acknowledgments}
We thank a number of people who read through a draft of this paper,
provided many helpful comments and highlighted issues which required clarity: Rob
Kennicutt, Lee Hartmann, Mark Heyer, Bruce Elmegreen, Fabian Heitsch,
Javier Ballesteros-Paredes. We thank Ralf Klessen for
providing a useful referee's report.  CLD also thanks Jin Koda for
providing the data for Figure 8, right panel, and Ian Bonnell for helpful discussions.
The research of A.B. is supported by a Max Planck Fellowship and by
the DFG Cluster of Excellence 
``Origin and Structure of the Universe''.
The calculations presented in this paper were primarily performed on
the HLRB-II: SGI Altix 4700 supercomputer and Linux cluster 
at the Leibniz supercomputer
centre, Garching. 
Run A was performed on the University of Exeter's SGI Altix ICE 8200 supercomputer.
Some of the figures in this paper were produced using \textsc{SPLASH}
\citep{splash2007}, a visualization tool for SPH that is publicly
available at http://www.astro.ex.ac.uk/people/dprice/splash.
\bibliographystyle{mn2e}
\bibliography{Dobbs}
\bsp
\label{lastpage}
\end{document}